\documentclass[twocolumn,trackchanges]{aastex701}

\begin{document}

\title{Differential Luminosity Response of Balmer-Line Profiles in Galactic B/A Luminous Stars}

\author[0000-0002-7449-0108]{N.~L.~Vaidman}
\affiliation{Fesenkov Astrophysical Institute, Observatory, 23, Almaty, 050020, Kazakhstan}
\affiliation{Al-Farabi Kazakh National University, Al-Farabi Ave., 71, 050040, Almaty, Kazakhstan}
\email{nva1dmann@gmail.com}


\author[0000-0001-5163-508X]{S.~A.~Khokhlov}
\affiliation{Fesenkov Astrophysical Institute, Observatory, 23, Almaty, 050020, Kazakhstan}
\affiliation{Al-Farabi Kazakh National University, Al-Farabi Ave., 71, 050040, Almaty, Kazakhstan}
\email{skhokh88@gmail.com}

\author[0009-0005-6851-7270]{Sh.~T.~Nurmakhametova}
\affiliation{Fesenkov Astrophysical Institute, Observatory, 23, Almaty, 050020, Kazakhstan}
\affiliation{Al-Farabi Kazakh National University, Al-Farabi Ave., 71, 050040, Almaty, Kazakhstan}
\email{shahidanurmahametova@gmail.com}

\author[0009-0007-1454-012X]{N.~A.~Litvinov}
\affiliation{Fesenkov Astrophysical Institute, Observatory, 23, Almaty, 050020, Kazakhstan}
\affiliation{Al-Farabi Kazakh National University, Al-Farabi Ave., 71, 050040, Almaty, Kazakhstan}
\email{litvinov@fai.kz}

\author[0000-0001-9788-7485]{A.~T. Agishev}
\affiliation{Fesenkov Astrophysical Institute, Observatory, 23, Almaty, 050020, Kazakhstan}
\affiliation{Al-Farabi Kazakh National University, Al-Farabi Ave., 71, 050040, Almaty, Kazakhstan}
\email{aldiyar.agishev@gmail.com}

\author[0000-0001-5437-8146]{D.~A. Alimbetova}
\affiliation{Al-Farabi Kazakh National University, Al-Farabi Ave., 71, 050040, Almaty, Kazakhstan}
\email{ada2030@mail.ru}

\begin{abstract}
We investigate the differential luminosity response of the optical Balmer series from H8 to H$\alpha$ in Galactic B5--A5 luminous stars using homogeneous measurements of central depth, FWHM, and equivalent width at $R\simeq12{,}000$. Direct line-to-line comparisons show a systematic change in the central-depth response toward the lower-order transitions. For the primary 72-star luminosity sample, the fitted luminosity coefficients show an overall positive trend with upper-level quantum number, with a point-estimate gradient of $dq/dn=0.0419$. A joint Monte Carlo analysis with 10,000 realisations gives a median $dq/dn=0.0387$ and a 95\% interval of $[0.0253,0.0532]$, with no non-positive gradients obtained. The higher-order H8--H$\delta$ coefficients form a statistically unresolved plateau, followed by a well-supported increase in response magnitude from H$\delta$ toward H$\alpha$. The result is robust to the adopted effective-temperature scale. The FWHM coefficients show no comparable Balmer-order progression, and the directly fitted profile-shape response is consistent with zero from H8 through H$\beta$. For H$\alpha$, the full logarithmic sample is not formally significant, whereas removal of the three remaining strongly filled-core profiles yields a significantly negative shape response. The observed behaviour is qualitatively compatible with the systematic variation of pressure broadening across the Balmer series, while the enhanced H$\alpha$ response may additionally reflect non-LTE and wind-related effects.
\end{abstract}


\section{Introduction}
\label{sec:introduction}

Blue supergiants are among the most luminous stars at optical wavelengths and represent a short-lived but important stage in massive-star evolution. Their spectra constrain post-main-sequence evolution, mass loss, rotational mixing, chemical enrichment, and stellar parameters, while their intrinsic brightness makes them accessible to quantitative spectroscopy at large Galactic and extragalactic distances. Late-B and A-type supergiants are especially useful because their optical spectra contain strong hydrogen lines together with numerous metal-line diagnostics \citep{2006A&A...445.1099P,2012A&A...543A..80F,2022A&A...668A..92W}.

The Balmer lines play a central role in these analyses. Their wings respond strongly to atmospheric density through Stark broadening and are therefore sensitive to surface gravity, while their strength and core morphology also depend on effective temperature. Consequently, Balmer profiles are interpreted together with independent temperature diagnostics rather than as one-parameter luminosity indicators. In the present work, luminosity is therefore treated as an external empirical stellar scale: any measured luminosity correlation may ultimately be mediated by gravity, atmospheric density structure, external broadening, non-LTE line formation, or winds.

The empirical connection between Balmer morphology and luminosity has a long history, from spectroscopic absolute magnitudes and the MK system \citep{1914ApJ....40..385A,1935ApJ....81..187A,1943assw.book.....M} to quantitative H$\gamma$ equivalent-width calibrations \citep{1964PDAO...12..317P,1974MNRAS.166..203B,2004MNRAS.353..601E}. The flux-weighted gravity--luminosity relationship provides a complementary physical framework \citep{2008ApJ...681..269K,2017AJ....154..102U}, while recent homogeneous surveys have introduced profile-based diagnostics such as ${\rm FW}_{3414}({\rm H}\beta)$ and documented the luminosity-class dependence of Balmer morphology \citep{2023A&A...674A.212D,2024A&A...690A.176N}. These studies establish that Balmer profiles contain luminosity information, but they generally treat individual transitions or atmospheric parameters rather than quantifying the differential response across several Balmer members and several profile observables.

Gaia astrometry and homogeneous spectroscopy now make such a differential comparison possible. Our previous work established a quality-controlled Gaia DR3/EDR3 distance scale for Galactic hot supergiants \citep{2025Univ...11..359V}; the same programme sample also has a homogeneous empirical effective-temperature scale. In the present study we use these stellar parameters together with homogeneous Balmer measurements to ask three related questions: whether the line-core response changes systematically from H8 to H$\alpha$, whether that internal sequence contains a component correlated with luminosity after controlling for $T_{\rm eff}$, and how the response is partitioned among central depth, FWHM, equivalent width, and profile shape.

The aim is not to introduce another single-line luminosity calibration. Instead, we characterize the optical Balmer series as a structured spectroscopic sequence and measure its differential, luminosity-correlated response in Galactic B5--A5 luminous stars. The contribution of the present work is therefore the homogeneous per-line, per-observable comparison across six Balmer transitions, together with explicit tests of temperature-scale dependence, sample composition, profile broadening, and line-shape systematics.

\section{Data and adopted stellar parameters}
\label{sec:data}

\subsection{Sample criteria and observational material}
\label{sec:sample}

Our target list consists of bright northern-hemisphere objects ($\delta \ge 30^{\circ}$) reaching down to $V \approx 9$ mag, selected as B5--A5 stars with luminosity classifications from bright giants (II) to supergiants (Ia, Iab, Ib). Initial spectral types were gathered from SIMBAD and the literature. Objects with discrepant published classifications were retained when at least one source indicated high luminosity; because luminosity-class II objects are included, we refer to the full ensemble as B5--A5 luminous stars.

The original programme contains 84 stars. Four objects do not enter the homogeneous TCO Balmer-line analysis because they either lack the required TCO Balmer coverage or are treated separately with dedicated multi-epoch material, leaving 80 stars with usable Balmer measurements. Eight of these have Gaia-based luminosities below $\log(L/L_{\odot})=3.5$ and are retained only for comparison. The primary luminosity-response sample therefore contains 72 stars.

The homogeneous spectra were obtained between 2011 and 2026 with the 0.81-m telescope at the Three College Observatory (TCO) using the fibre-fed Shelyak Instruments eShel spectrograph.\footnote{\url{https://www.shelyak.com}} The nominal resolving power is $R\simeq12{,}000$. The post-2018 configuration provides blue coverage to approximately 3800\,\AA, while spectra obtained before the autumn-2018 fibre replacement begin near 4200\,\AA; the red limit is approximately 7900\,\AA. Consequently, H8, H$\epsilon$, and most H$\delta$ measurements are drawn predominantly from the post-2018 material. A homogeneous epoch-by-epoch reconstruction of the line-spread function is not available; possible configuration dependence is therefore tested empirically by repeating the analysis on the post-2018 subset in Section~\ref{sec:luminosity_robustness}. The median local S/N estimated by MERLIS is approximately 10 near H8, 33 near H$\epsilon$, 17 near H$\delta$, and 50 near H$\gamma$, while H$\beta$ and H$\alpha$ typically exceed S/N$\simeq100$.

Although multiple epochs are available for several targets, the statistical analysis uses one adopted spectrum per star so that objects with extensive time series do not receive disproportionate weight. The available repeated observations are used separately as an empirical repeatability check in Section~\ref{sec:luminosity_robustness}.

Wavelength calibration was performed using ThAr comparison spectra, with approximately 800--1000 identified lines and a typical wavelength-solution rms of $\sim0.03\,\AA$.

\subsection{Data reduction and digital preservation}
\label{sec:reduction}

TCO frames were processed with the standard IRAF \texttt{echelle} procedures for bias removal, spectral extraction, wavelength calibration, and frame combination. Flat-field correction was not applied, following the established TCO reduction procedure \citep{2023Galax..11....8M}. The detector pixel-to-pixel sensitivity variation is $\lesssim1.5\%$. For a broad Balmer profile sampled by roughly twenty detector pixels across a typical FWHM, an uncorrelated 1.5\% pixel response would average to approximately 0.3\% in the fitted central depth, smaller than the typical depth uncertainties of about 1.5--3\%. This estimate is used only as a quantitative check on the expected scale of the omitted flat-field correction.

All Balmer-line parameters were measured homogeneously with the MERLIS pipeline \citep{2026ApJ...1008....4V}. For each usable transition, the analysis employs the equivalent width ($EW$), central depth ($d$), and FWHM. The detailed profile fitting, local continuum treatment, measurement boundaries, and uncertainty estimation are described in \citet{2026ApJ...1008....4V}.

The fully reduced, continuum-normalized spectral library is publicly available through the Kazakhstan National Virtual Observatory (KazVO)\footnote{\url{https://vo.fai.kz/}}, with ObsCore metadata and SSAP access.

\subsection{Effective temperatures and luminosities}
\label{sec:parameters}

Effective temperatures were adopted from our preceding empirical calibration study of Galactic B5--A5 luminous stars \citep{vaidman2026empiricaleffectivetemperaturecalibrationsgalactic}. The adopted scale combines 42 optical spectroscopic calibrations. Three contain Balmer central depths, $d({\rm H}\epsilon)$, $d({\rm H}\delta)$, and $d({\rm H}\gamma)$, and together contribute approximately 9\% of the total inverse-variance weight. Balmer equivalent widths, FWHM measurements, and $d({\rm H}\beta)$ do not enter the temperature estimator.

The stellar luminosities were determined in the companion analysis of \citet{galaxies14050087}. The absolute visual magnitude was calculated as

\begin{equation}
\mathbf{
M_V
=
m_V
-
5\log\left(\frac{d}{10\,{\rm pc}}\right)
-
A_V ,
}
\end{equation}

where $m_V$ is the adopted apparent magnitude, $d$ is the astrometric distance, and $A_V$ is the line-of-sight extinction. The distance scale follows the quality-controlled Gaia DR3 analysis of \citet{2025Univ...11..359V}. Reddening was obtained from the Bayestar19 three-dimensional dust map \citep{2019ApJ...887...93G}, evaluated at the stellar coordinates and adopted distance; its uncertainty was estimated from the reddening values at $d-\sigma_d$ and $d+\sigma_d$. We adopted $A_V=3.1E(B-V)$ \citep{1989ApJ...345..245C} and interpolated $BC_V$ from \citet{1998A&A...333..231B}.

The luminosity uncertainty was propagated from distance and extinction as

\begin{equation}
\mathbf{
\sigma_{\log L}
=
\left[
\left(
\frac{2}{\ln 10}\frac{\sigma_d}{d}
\right)^2
+
\left(
0.4\,\sigma_{A_V}
\right)^2
\right]^{1/2}.
}
\end{equation}

The uncertainties associated with apparent magnitude and bolometric correction are negligible compared with these terms for the present sample. The complete per-star values of $m_V$, $A_V$, $\sigma_{A_V}$, $BC_V$, distance, $T_{\rm eff}$, $\log(L/L_{\odot})$, their uncertainties, the Balmer measurements, and the quality, multiplicity, and variability flags are supplied in the machine-readable table archived at \href{https://doi.org/10.5281/zenodo.23008571}{doi:10.5281/zenodo.23008571}.

Multiplicity and photometric variability were considered explicitly. A WDS companion is reported for 17 of the 72 primary-sample stars. The estimated unresolved-companion luminosity bias, $\log(1+F_2/F_1)$, has a median of 0.000 dex, a 90th percentile of 0.008 dex, and a maximum of 0.056 dex. Excluding the three potentially blended close pairs gives $dq/dn=0.0424$, while excluding all WDS-companion stars gives $dq/dn=0.0419$, identical to the full-sample point estimate.

VSX variability information is available for 23 of the 72 stars, including 12 $\alpha$~Cyg variables. After screening inconsistent catalogue entries, the median peak-to-peak amplitude is 0.070 mag, corresponding to $\sigma(\log L)=0.0099$ dex, compared with a median adopted luminosity uncertainty of 0.070 dex. The variability amplitude correlates with luminosity ($\rho_{\rm S}=+0.42$, $p=0.044$), so single-epoch variability is not assumed to be luminosity independent.

\begin{figure}
\centering
\includegraphics[width=0.5\textwidth,height=0.35\textheight,keepaspectratio]{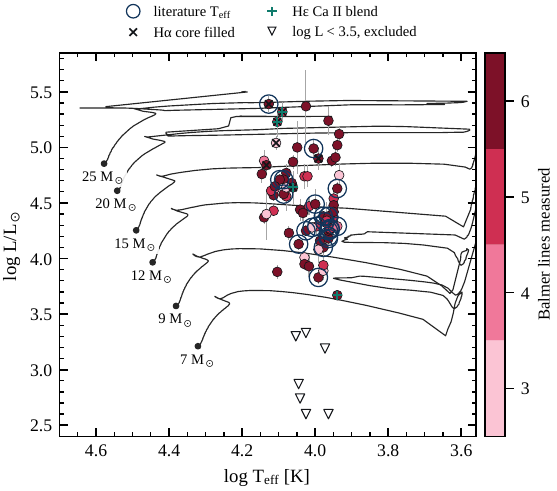}
\caption{Distribution of the analysed stars in the $T_{\rm eff}$--$\log(L/L_{\odot})$ plane. The colour scale gives the number of Balmer transitions with usable measurements. Open circles identify objects with independent literature $T_{\rm eff}$ determinations, red crosses mark strongly filled H$\alpha$ cores, and orange plus symbols mark the H$\epsilon$/Ca\,{\sc ii}~H contamination flag. Grey inverted triangles show the eight objects with $\log(L/L_{\odot})<3.5$, excluded from the primary luminosity-response sample. Solid curves show rotating Geneva tracks for $Z=0.014$ and $v_{\rm ini}/v_{\rm crit}=0.4$ \citep{2012A&A...537A.146E}; they are included for context only.}
\label{fig:sample_hrd}
\end{figure}

For the 72-star primary sample, $T_{\rm eff}$ and luminosity are only modestly correlated, with $\rho_{\rm S}=+0.25$ and $p=0.036$. Figure~\ref{fig:sample_hrd} shows the sample together with rotating Geneva tracks for $Z=0.014$ and $v_{\rm ini}/v_{\rm crit}=0.4$ \citep{2012A&A...537A.146E}. The tracks are shown only for evolutionary context and are not used in the Balmer regressions.

Because three empirical temperature calibrations contain Balmer central depths, the possible circularity of the adopted $T_{\rm eff}$ scale was tested explicitly by removing those calibrations and, independently, by substituting published temperatures from quantitative atmospheric analyses. The procedures are described in Section~\ref{sec:robustness_method} and the results in Sections~\ref{sec:luminosity_sequence}--\ref{sec:luminosity_robustness}.

\section{Method}
\label{sec:method}

\subsection{Balmer-series diagnostics}
\label{sec:line_selection}

The analysis includes H8, H$\epsilon$, H$\delta$, H$\gamma$, H$\beta$, and H$\alpha$. For each transition we consider central depth, equivalent width, and FWHM, with central depth used as the primary diagnostic of the differential Balmer-series behaviour.

H8 at $\lambda3889.05$ is blended with the nearly coincident He\,{\sc i} $\lambda3888.65$ transition. At $R\simeq12{,}000$ and for the intrinsic widths of these stellar profiles, the two contributions are not resolved as separate components, and the He\,{\sc i} contribution varies across the B5--A5 temperature range. H$\epsilon$ is affected by Ca\,{\sc ii}~H near $\lambda3968$, while H$\alpha$ can be strongly modified by emission filling and winds.

Potentially compromised measurements are flagged rather than removed from the baseline fit. Enhanced H$\epsilon$ absorption is flagged when $d({\rm H}\epsilon)-d({\rm H}\gamma)>0.12$, identifying four stars, and strongly filled H$\alpha$ cores are flagged when $d({\rm H}\alpha)<0.06$, identifying seven stars in the parent sample. An anomalous H8 measurement for HD~12301 was identified during the quality-control analysis and is treated as a separate quality flag. This measurement is not included in the H8 relation shown in Figure~\ref{fig:depth_sequence}. The effect of removing flagged measurements is quantified in Section~\ref{sec:luminosity_robustness}.

\subsection{Internal Balmer core-depth relations}
\label{sec:depth_method}

To characterize the series independently of the adopted stellar parameters, the central depth of each transition was compared with H$\gamma$, chosen as a well-measured reference line with fewer endpoint complications. For each Balmer member we fitted

\begin{equation}
d({\rm H}n)
=
a_n
+
b_n\,d({\rm H}\gamma),
\end{equation}

where $b_n$ describes the relative dynamic response of the line depth over the observed range. H$\gamma$ has $b=1$ by construction and is excluded from the trend fit in $b_n$ versus upper-level quantum number $n$.

The relations were fitted with an errors-in-variables maximum-likelihood formulation that includes measurement uncertainties in both depths and an additional intrinsic-scatter term. The inferred intrinsic scatter is typically $0.03$--$0.09$ in normalized depth, compared with individual measurement uncertainties of approximately $0.01$--$0.02$. Pearson's $r$ and the residual scatter characterize each pairwise relation; the Balmer-order dependence is summarized by $db/dn$ and Spearman's $\rho_{\rm S}$. The analysis was repeated with H$\delta$ and H$\beta$ as alternative reference transitions and was also applied to the equivalent widths and FWHM measurements.

\subsection{Luminosity response of the Balmer-line cores}
\label{sec:luminosity_method}

For each Balmer line, the central depth was represented by

\begin{equation}
d({\rm H}n)
=
A_n
+
C_{1,n}\,\tau
+
C_{2,n}\,\tau^{2}
+
q_n\,\ell ,
\label{eq:luminosity_model}
\end{equation}

with

\begin{equation}
\tau=\frac{T_{\rm eff}}{10^{3}\ {\rm K}}-11
\end{equation}

and

\begin{equation}
\ell=\log\left(\frac{L}{L_{\odot}}\right).
\end{equation}

The coefficient $q_n$ measures the change in central depth per dex in luminosity at fixed $T_{\rm eff}$; negative $q_n$ corresponds to a shallower core at higher luminosity. The centered temperature coordinate $\tau$ is a numerical reparameterization of the same quadratic temperature dependence and does not alter $q_n$. It avoids the near-collinearity between an uncentered temperature and its square: in the fitted basis ${\rm corr}(\tau,\tau^2)=+0.11$, and the variance-inflation factors for $\tau$, $\tau^2$, and $\ell$ are 1.10, 1.03, and 1.11, respectively.

The adopted quadratic form was tested against first- and third-order temperature descriptions, and the fitted residuals were examined for remaining correlation with $T_{\rm eff}$. The luminosity coefficients were then fitted as

\begin{equation}
q_n
=
q_0
+
\frac{dq}{dn}\,n .
\end{equation}

For each transition, the coefficients in Equation~(\ref{eq:luminosity_model}) were obtained by unweighted ordinary least-squares regression over the available stars. The baseline point estimates therefore do not use the individual measurement uncertainties as regression weights; instead, uncertainties in central depth, $T_{\rm eff}$, and luminosity are propagated through the joint Monte Carlo procedure described below.

Uncertainties in the sequence were evaluated with 10,000 joint Monte Carlo realisations. In each realisation the stars were resampled with replacement and the measured central depths, $T_{\rm eff}$, and $\log(L/L_{\odot})$ were perturbed according to their adopted uncertainties. The covariance between $T_{\rm eff}$ and luminosity introduced through the temperature-dependent bolometric correction was included. The same resampled and perturbed stellar sample was used for all six transitions, preserving covariance among the resulting $q_n$ coefficients when $dq/dn$ was fitted.

\subsection{Robustness tests}
\label{sec:robustness_method}

The principal analysis was repeated under several controlled changes. We tested individual influence by removing each star in turn; sample composition by restricting to stars with all six Balmer depths; line leverage by excluding each transition in turn; and profile quality by removing the objective H$\epsilon$ and H$\alpha$ flags, applying Huber robust regression, and restricting to MERLIS measurements with status ``detected''.

Temperature-scale sensitivity was tested by (1) recomputing $T_{\rm eff}$ after removing the three Balmer-depth calibrations, (2) repeating the analysis for the 22 stars with independent literature temperatures using the same stellar subset for both temperature scales, and (3) imposing an artificial luminosity-correlated bias $T_{\rm eff}\rightarrow T_{\rm eff}+g[\log(L/L_{\odot})-4.5]$ with $g$ from $-3000$ to $+3000$~K~dex$^{-1}$.

The possible contribution of external line broadening was tested with an empirical $v_{\rm broad}$ derived from Mg\,{\sc ii} $\lambda4481$ and Si\,{\sc ii} $\lambda6347$. The instrumental contribution corresponding to $R\simeq12{,}000$ ($\simeq25$~km~s$^{-1}$) was removed in quadrature. At this resolution, $v_{\rm broad}$ is treated as a combined non-thermal broadening measure rather than a decomposition into rotation and macroturbulence. The Balmer regressions were repeated with $v_{\rm broad}$ as an additional covariate, and a controlled convolution experiment estimated the differential response expected from the measured broadening--luminosity relation alone.

Instrumental sensitivity was assessed by repeating the analysis using only the post-2018 fibre configuration. Multi-epoch spectra of HD~40297 and $\alpha$~Cyg were used as empirical repeatability checks; because these time series include intrinsic stellar variability, they are interpreted as upper bounds on repeatability scatter rather than pure measurement-error estimates.

\subsection{Depth, width, and equivalent-width response}
\label{sec:profile_response}

For each Balmer transition and each observable $X\in\{d,\mathrm{FWHM},\mathrm{EW}\}$, we fitted the logarithm directly,

\begin{equation}
\ln X
=
A_X
+
C_{1,X}\,\tau
+
C_{2,X}\,\tau^2
+
s_X\,\ell ,
\end{equation}

where $\tau=T_{\rm eff}/(10^3\,{\rm K})-11$ and

\begin{equation}
s_X
=
\left.
\frac{\partial \ln X}{\partial \ell}
\right|_{T_{\rm eff}}
\end{equation}

is the fractional luminosity response per dex at fixed effective temperature. No normalization by an adopted reference value is required.

To quantify profile shape, we define

\begin{equation}
\kappa
=
\frac{\mathrm{EW}}
{d\,\mathrm{FWHM}},
\end{equation}

so that

\begin{equation}
\frac{\partial \ln \kappa}{\partial \ell}
=
s_{\mathrm{EW}}
-
s_d
-
s_{\mathrm{FWHM}} .
\label{eq:kappa_response}
\end{equation}

Because $d$, FWHM, and $EW$ come from the same locally normalized spectrum, their errors are not independent. The current MERLIS output does not provide the full per-spectrum covariance matrix, so the marginal errors are not propagated as if they were independent. Instead, $\kappa$ is constructed for each spectrum and fitted directly as

\begin{equation}
\ln\kappa
=
A_{\kappa}
+
C_{1,\kappa}\,\tau
+
C_{2,\kappa}\,\tau^2
+
s_{\kappa}\,\ell .
\end{equation}

Whole-star bootstrap resampling preserves the observed sample-level joint variation of $d$, FWHM, and $EW$, although it does not reconstruct the full within-spectrum measurement covariance. Equation~(\ref{eq:kappa_response}) is therefore used only as an algebraic identity for interpretation.

\section{Results}
\label{sec:results}

\subsection{A systematic sequence in Balmer-line core depths}
\label{sec:depth_sequence}

\begin{figure*}[t]
\centering
\includegraphics[width=\textwidth]{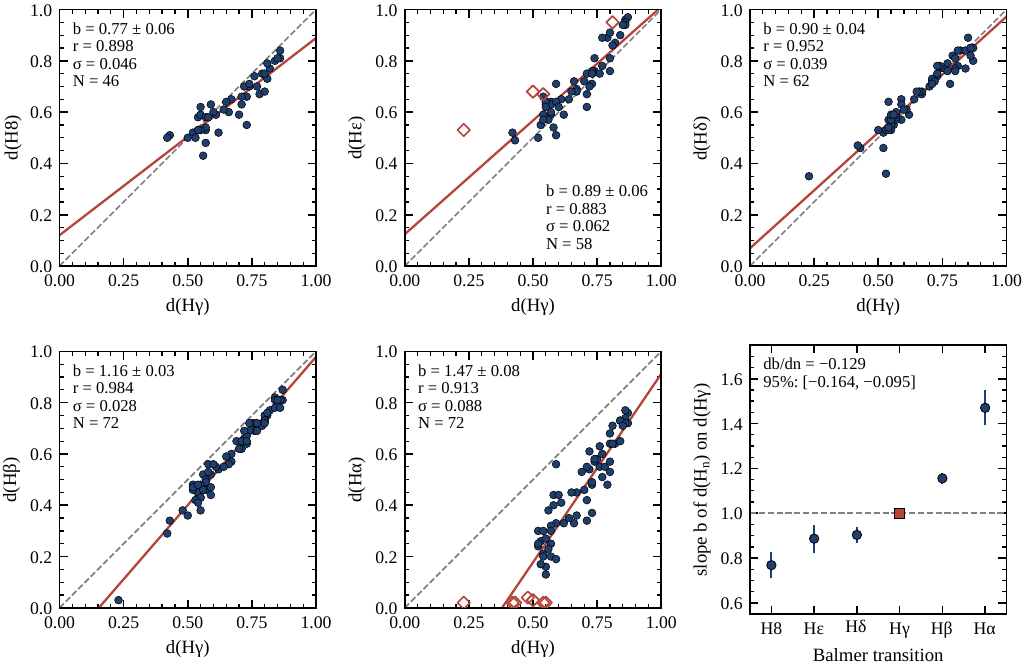}
\caption{
Central-depth relations across the optical Balmer series.
The first five panels compare the central depths of H8, H$\epsilon$,
H$\delta$, H$\beta$, and H$\alpha$ with that of H$\gamma$, adopted as
the reference transition. Dashed lines indicate the one-to-one relation,
while solid lines show the corresponding linear fits. Red symbols mark measurements affected by identified profile complications: enhanced H$\epsilon$ absorption associated with blending by Ca\,{\sc ii}~H and strongly filled H$\alpha$ cores.
The lower-right panel shows the fitted slopes, $b$, as a function of the
upper-level principal quantum number, $n$. H$\gamma$ is fixed at $b=1$
by construction and is shown only as a reference point; it is not included
in the fit to the slope sequence. The fitted slopes exhibit a systematic
increase toward the lower-order Balmer transitions.
}
\label{fig:depth_sequence}
\end{figure*}

We first examined the internal behaviour of the Balmer series without reference to the adopted $T_{\rm eff}$ or luminosity scales. For each transition, the central line depth was compared with that of H$\gamma$, adopted as the reference member of the series. The resulting relations are shown in Figure~\ref{fig:depth_sequence}.

The Balmer-line depths are strongly correlated, but their relations with H$\gamma$ depart systematically from one-to-one scaling. The fitted slope increases continuously toward the lower-order members of the series: $b=0.77\pm0.06$ for H8, $0.89\pm0.06$ for H$\epsilon$, and $0.90\pm0.04$ for H$\delta$, followed by the reference value $b=1$ for H$\gamma$, $1.16\pm0.03$ for H$\beta$, and $1.47\pm0.08$ for H$\alpha$.

The H$\beta$--H$\gamma$ relation is particularly noteworthy. It is the tightest relation in the set, with $r=0.984$ and a residual scatter of only $\sigma=0.028$, yet its slope is clearly larger than unity. The enhanced response of the H$\beta$ core relative to H$\gamma$ is therefore not associated with increased scatter. H$\delta$ also follows H$\gamma$ closely ($r=0.952$, $\sigma=0.039$), but with a slope slightly below unity.

Larger departures from the main relations occur toward the ends of the observed series. H8 remains well correlated with H$\gamma$ ($r=0.898$, $\sigma=0.046$). During the quality-control analysis, HD~12301 was identified as having an anomalously shallow H8 core relative to H$\gamma$; this measurement is not included in the H8 relation shown in Figure~\ref{fig:depth_sequence}, and the object has no adopted luminosity and therefore does not enter the luminosity-response regressions. In addition, H8 is blended with the nearly coincident He\,{\sc i} $\lambda3888.65$ transition, whose contribution varies with temperature across the B5--A5 range, making H8 less suitable as a purely hydrogen-line diagnostic. H$\epsilon$ gives $r=0.883$ and $\sigma=0.062$, with several objects displaying systematically enhanced depths consistent with additional absorption from the neighbouring Ca\,{\sc ii}~H $\lambda3968$ feature.

H$\alpha$ displays a qualitatively different departure. A subset of stars has strongly reduced or nearly filled H$\alpha$ cores despite substantial H$\gamma$ absorption, consistent with emission filling in wind-affected profiles. These objects form a distinct low-depth branch in Figure~\ref{fig:depth_sequence}. The principal H$\alpha$ relation nevertheless remains well defined, indicating that the large H$\alpha$ slope is not produced solely by these extreme profiles.

The fitted slope point estimates form a monotonic sequence with upper-level quantum number $n$. Excluding H$\gamma$, whose value $b=1$ is fixed by definition, the dependence is characterized by

\begin{equation}
\frac{db}{dn}
=
-0.129 .
\end{equation}

The fitted gradient has a 95\% interval of $[-0.164,-0.095]$. The point estimates are fully monotonic, with $\rho_{\rm S}=-1.00$ and an exact $p$-value of $0.017$, although the adjacent H$\epsilon$ and H$\delta$ coefficients are not individually resolved within their uncertainties. Thus, the relative response of the Balmer-line cores changes systematically across the optical series, becoming progressively stronger toward the lower-order transitions.

The ordering is not an artefact of the choice of H$\gamma$ as the reference transition. Repeating the line-to-line analysis with H$\delta$ and H$\beta$ as alternative anchors preserves the monotonic ordering of the fitted slope point estimates, with $\rho_{\rm S}=-1.00$ in all three cases. Thus, the inferred ordering is preserved when the reference transition is changed.

An analogous but markedly weaker ordering is present in the equivalent widths. The fitted EW slopes are 1.07, 1.07, 1.13, 1.19, and 1.35 for H8, H$\epsilon$, H$\delta$, H$\beta$, and H$\alpha$, respectively, giving a Balmer-order gradient of $db_{\rm EW}/dn=-0.052\pm0.012$ with $\rho_{\rm S}=-0.90$. This gradient is about 2.5 times shallower than the corresponding central-depth gradient, $db_d/dn=-0.129\pm0.023$. In contrast, the FWHM slopes show no comparable ordering, with $db_{\rm FWHM}/dn=+0.064\pm0.038$ and $\rho_{\rm S}=+0.70$. The internal progression is therefore strongest in the line cores, while the integrated absorption shows the same trend in substantially diluted form, consistent with the absence of a comparable Balmer-order progression in the FWHM slopes.

\subsection{Luminosity-correlated response along the Balmer series}
\label{sec:luminosity_sequence}

The ordered line-to-line behaviour demonstrates that the Balmer cores do not vary identically across the series. To determine whether this differential response is associated with stellar luminosity, we next quantified the luminosity dependence of each central depth while explicitly controlling for $T_{\rm eff}$.

\begin{figure*}[t]
\centering
\includegraphics[width=\textwidth]{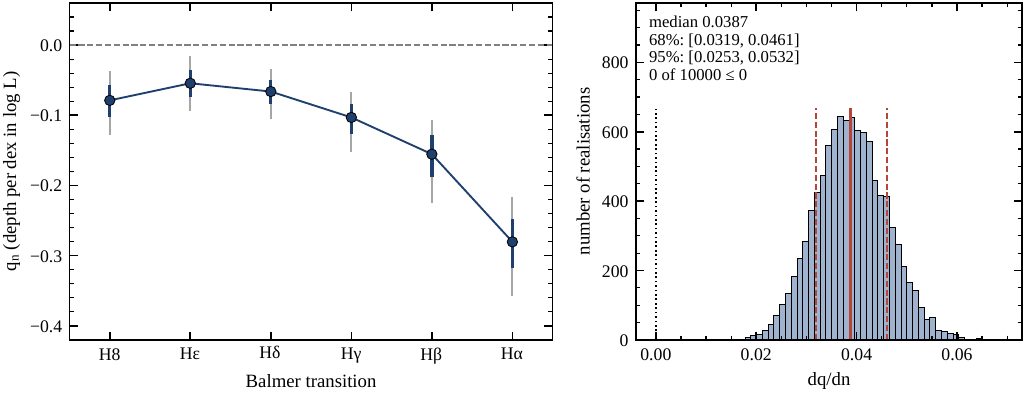}
\caption{
Luminosity-correlated response of the Balmer-line central depths after
accounting for the effective-temperature dependence.
\textit{Left:} Luminosity coefficient, $q_n$, for each Balmer transition
as a function of the upper-level principal quantum number, $n$.
Negative values of $q_n$ indicate decreasing central depth with increasing
luminosity. The dotted line shows the fitted dependence of $q_n$ on $n$.
\textit{Right:} Monte Carlo distribution of the corresponding gradient,
$dq/dn$, obtained from 10,000 realisations that propagate the uncertainties
in central depth, $T_{\rm eff}$, and luminosity while preserving the covariance
among the six $q_n$ coefficients.}
The solid vertical line marks the median of the distribution,
and the dashed lines delimit the 95\% interval. The interval
lies entirely above zero, indicating that the luminosity-correlated
response becomes systematically stronger toward the lower-order members
of the Balmer series.
\label{fig:luminosity_sequence}
\end{figure*}

For each transition, the coefficient $q_n$ measures the change in central line depth per dex in $\log(L/L_{\odot})$ at fixed effective temperature. The resulting coefficients are shown in Figure~\ref{fig:luminosity_sequence}. All six coefficients are negative, demonstrating that the Balmer-line cores become systematically shallower with increasing luminosity. Moreover, the magnitude of this response increases toward the lower-order members of the series.

The luminosity dependence changes only moderately among H8, H$\epsilon$, and H$\delta$, and becomes progressively stronger toward H$\gamma$, H$\beta$, and H$\alpha$. For the revised 72-star sample, a linear fit of $q_n$ as a function of upper-level quantum number gives

\begin{equation}
\frac{dq}{dn}
=
0.0419
\ {\rm dex}^{-1}\ {\rm per\ unit}\ n .
\end{equation}

The coefficients remain positively correlated with $n$, with $\rho_{\rm S}=+0.83$ and $p=0.042$. The joint Monte Carlo analysis gives a median gradient of $dq/dn=0.0387$, with 68\% and 95\% intervals of $[0.0319,0.0461]$ and $[0.0253,0.0532]$, respectively. None of the 10,000 realisations yields a non-positive gradient.

The joint Monte Carlo realisations also show that the individual adjacent coefficients are not all statistically separated. The probabilities that the absolute luminosity response increases between successive transitions are 0.18 for H8$\rightarrow$H$\epsilon$, 0.72 for H$\epsilon\rightarrow$H$\delta$, 0.998 for H$\delta\rightarrow$H$\gamma$, and effectively unity for H$\gamma\rightarrow$H$\beta$ and H$\beta\rightarrow$H$\alpha$. The probability that all five adjacent inequalities hold simultaneously is 0.072. We therefore describe the higher-order H8--H$\delta$ coefficients as a statistically unresolved plateau, followed by a well-supported increase in the magnitude of the luminosity response from H$\delta$ toward H$\alpha$.

We further tested whether the inferred progression could be introduced by the adopted effective-temperature scale. Recomputing $T_{\rm eff}$ using all 42 empirical calibrations gives $dq/dn=0.0421$, whereas excluding the three calibrations involving Balmer-line depths gives $dq/dn=0.0416$. A paired bootstrap gives $\Delta(dq/dn)=-0.0005$, with a 95\% interval of $[-0.0013,0.0000]$. Thus, removing the Balmer contribution to the temperature estimator changes the differential gradient by only about 1\%.

As an independent check, we repeated the analysis for the 22 stars with published $T_{\rm eff}$ determinations from quantitative atmospheric analyses. Using a linear temperature term for this smaller subset, the literature temperatures give $dq/dn=0.0406$, with a bootstrap 68\% interval of $[0.032,0.048]$, while our adopted temperatures give $dq/dn=0.0392$ for exactly the same stars; only 0.3\% of the literature-temperature bootstrap realisations yield a non-positive gradient. Retaining a quadratic temperature term gives corresponding gradients of $0.0353$ and $0.0340$. The close agreement between the two temperature scales under either specification indicates that the measured progression is not produced by the empirical $T_{\rm eff}$ calibration.

The luminosity-correlated response therefore changes systematically along the Balmer series, although the adjacent coefficients are not all individually resolved. At fixed $T_{\rm eff}$, the higher-order H8--H$\delta$ transitions show the weakest and statistically similar core-depth responses, whereas the magnitude of the response increases from H$\delta$ through H$\gamma$ and H$\beta$ to H$\alpha$.

\subsection{Robustness of the luminosity ordering}
\label{sec:luminosity_robustness}

No individual star controls the inferred gradient. In a leave-one-star-out analysis, repeating the complete fit after removing each of the 72 stars in turn gives $dq/dn$ values between 0.0394 and 0.0470.

The result is also insensitive to the adopted polynomial order in temperature. Linear, quadratic, and cubic temperature descriptions give $dq/dn=0.0396\pm0.0119$, $0.0419\pm0.0127$, and $0.0402\pm0.0126$, respectively, with $\rho_{\rm S}=+0.83$ in all three cases. For the adopted quadratic model, the residual correlations with $T_{\rm eff}$ range from $\rho_{\rm S}=-0.09$ to $+0.04$ across the six lines, with $p\geq0.52$ throughout.

As a deliberately conservative stress test, we imposed an artificial luminosity-correlated temperature bias, $T_{\rm eff}\rightarrow T_{\rm eff}+g[\log(L/L_{\odot})-4.5]$, with $g$ ranging from $-3000$ to $+3000$~K~dex$^{-1}$. Across this full range, the Balmer-series gradient remains positive, varying from $dq/dn=0.0324$ to $0.0534$, compared with the baseline value of $0.0419$.

Differences in line availability likewise do not generate the progression. All six central depths are available for 44 of the 72 primary-sample stars. Restricting the analysis to this common subset gives $dq/dn=0.0351\pm0.0131$ with $\rho_{\rm S}=+0.83$; the 68\% bootstrap interval is $[0.0275,0.0433]$, and none of 2000 bootstrap realisations yields a non-positive gradient. The largest change in an individual coefficient is for H$\gamma$, from $-0.112$ to $-0.091$, smaller than its quoted uncertainty.

Objective profile flags and robust fitting produce the same qualitative result. The baseline gradient is $dq/dn=0.0419$; removing the flagged H$\epsilon$ and H$\alpha$ measurements gives $0.0348$, Huber regression gives $0.0396$, Huber regression after flag removal gives $0.0337$, and restricting the analysis to MERLIS measurements with status "detected" gives $0.0402$. In all cases $\rho_{\rm S}=+0.83$. Thus, the amplitude of the fitted gradient changes moderately under the most restrictive selections, but the Balmer-order dependence is preserved.

The strongest individual luminosity response is observed for H$\alpha$, where wind emission can substantially modify the line core. We therefore tested whether H$\alpha$ alone determines the inferred sequence by repeating the fit after removing each Balmer transition in turn. The resulting gradients are summarized in Table~\ref{tab:jackknife}.

\begin{deluxetable}{lccc}
\tabletypesize{\scriptsize}
\tablewidth{0pt}
\tablecaption{Leave-One-Line-Out Tests of the Luminosity-Gradient Sequence
\label{tab:jackknife}}
\tablehead{
\colhead{Excluded Line} &
\colhead{$dq/dn$} &
\colhead{$\rho_{\rm S}$} &
\colhead{$N_{\rm lines}$}
}
\startdata
None        & $0.0419 \pm 0.0127$ & $+0.83$ & 6 \\
H8          & $0.0583 \pm 0.0129$ & $+1.00$ & 5 \\
H$\epsilon$ & $0.0405 \pm 0.0158$ & $+0.90$ & 5 \\
H$\delta$   & $0.0405 \pm 0.0134$ & $+0.90$ & 5 \\
H$\gamma$   & $0.0433 \pm 0.0134$ & $+0.70$ & 5 \\
H$\beta$    & $0.0451 \pm 0.0152$ & $+0.70$ & 5 \\
H$\alpha$   & $0.0226 \pm 0.0093$ & $+0.70$ & 5
\enddata
\tablecomments{
The gradient was recomputed from the current six $q_n$ point estimates after excluding each transition in turn. The quoted uncertainty is the standard error of the linear slope across the remaining $q_n$ values. The first row gives the complete six-line sequence.
}
\end{deluxetable}

Excluding any one transition leaves a positive point-estimate gradient. Removing H$\alpha$ gives $dq/dn=0.0226\pm0.0093$, whereas removing H8 gives $0.0583\pm0.0129$ and $\rho_{\rm S}=+1.00$. The progression is therefore weakened, but not eliminated, when the wind-sensitive H$\alpha$ endpoint is removed.

The predominantly photospheric H$\delta$--H$\gamma$--H$\beta$ subset shows the same ordering, with $q=-0.071$, $-0.112$, and $-0.168$, respectively, and $dq/dn=0.0485\pm0.0043$. Because only three transitions enter this fit, its formal uncertainty is descriptive rather than an independent significance estimate.

The 2018 fibre change does not account for the result. Restricting the analysis to the 59 stars observed in the post-2018 configuration gives $dq/dn=0.0400$ with $\rho_{\rm S}=+0.83$, compared with $0.0419$ for the full sample.

Repeated spectra provide an empirical check on repeatability and intrinsic variability. For HD~40297, three epochs spanning 2015--2021 show essentially unchanged H$\gamma$ and H$\beta$ central depths and an H$\alpha$ depth range of 0.010 (about 2\%). In 14 spectra of $\alpha$~Cyg, the relative rms scatter in the central depth increases from approximately 7\% at H$\gamma$ to 15\% at H$\beta$ and 64\% at H$\alpha$. These time series include genuine stellar variability and are therefore treated as upper bounds on measurement repeatability rather than as pure instrumental errors.

External broadening provides an additional potential contribution to the luminosity-correlated variation of the Balmer-line central depths. To test this effect independently of the pressure-broadened hydrogen profiles, we estimated an empirical broadening velocity, $v_{\rm broad}$, from the Mg\,{\sc ii} $\lambda4481$ and Si\,{\sc ii} $\lambda6347$ metal lines after removing the instrumental contribution corresponding to $R\simeq12{,}000$ ($\simeq25$~km~s$^{-1}$) in quadrature. At the present spectral resolution, $v_{\rm broad}$ is treated as an empirical measure of the combined non-thermal line broadening, and we do not attempt to separate the rotational and macroturbulent contributions.

The resulting broadening measure has a median value of 57~km~s$^{-1}$, with a 16--84 percentile range of 43--71~km~s$^{-1}$. The estimates obtained independently from Mg\,{\sc ii} and Si\,{\sc ii} agree closely, with $\rho_{\rm S}=+0.93$ and no median offset. Where published $v\sin i$ measurements are available, the metal-line broadening estimates also correlate positively with them, with $\rho_{\rm S}=+0.57$ for Mg\,{\sc ii} ($N=29$) and $\rho_{\rm S}=+0.53$ for Si\,{\sc ii} ($N=43$).

The metal-line broadening increases with luminosity at $dv_{\rm broad}/d\log(L/L_{\odot})=13\pm4$~km~s$^{-1}$~dex$^{-1}$. The corresponding slopes obtained from the individual lines are $10\pm3$~km~s$^{-1}$~dex$^{-1}$ for Mg\,{\sc ii} and $15\pm5$~km~s$^{-1}$~dex$^{-1}$ for Si\,{\sc ii}. The correlation remains significant after controlling for effective temperature, with a partial Spearman coefficient of $\rho=+0.54$ and $p<10^{-4}$.

When $v_{\rm broad}$ is included as an additional covariate in the Balmer-depth regressions, the differential Balmer-series gradient decreases from $dq/dn=0.0419$ to $0.0368$, corresponding to a reduction of approximately 12\%, while the rank correlation remains $\rho_{\rm S}=+0.83$. Bootstrap resampling gives a 68\% interval of $[0.027,0.044]$, with only 0.05\% of the realisations yielding a non-positive gradient. The measured luminosity dependence of external broadening therefore contributes to the observed differential response but does not remove the Balmer-series progression.

A complementary convolution experiment gives a consistent result. When evaluated at the measured broadening--luminosity slope of 13~km~s$^{-1}$~dex$^{-1}$, external broadening reproduces approximately 16\% of the observed differential gradient. A broadening--luminosity slope of approximately 80~km~s$^{-1}$~dex$^{-1}$ would be required for external broadening alone to reproduce the full observed sequence. The two tests therefore indicate that luminosity-correlated external broadening accounts for approximately 12-16\% of the measured differential response.

For comparison, ${\rm FWHM}({\rm H}\gamma)$ is anticorrelated with the metal-line broadening measure, with $\rho_{\rm S}=-0.68$. This confirms that the H$\gamma$ width contains a strong pressure-broadening contribution and is not an appropriate empirical proxy for rotational or macroturbulent broadening. We therefore do not use ${\rm FWHM}({\rm H}\gamma)$ as an external-broadening correction.

\subsection{Depth, width, and profile-shape response}
\label{sec:profile_response_results}

We next compared the luminosity-correlated responses of the central depth, FWHM, and equivalent width using the direct logarithmic regressions defined in Section~\ref{sec:profile_response}. The resulting coefficients therefore represent $\partial\ln X/\partial\log(L/L_{\odot})$ at fixed $T_{\rm eff}$ and do not require normalization by a reference value of the observable. The results are summarized in Table~\ref{tab:profile_response}.

The central-depth response becomes progressively stronger toward the lower-order Balmer transitions. A linear fit to the logarithmic depth coefficients as a function of upper-level quantum number gives a series gradient of $+0.154\pm0.048$ per unit $n$. In contrast, the FWHM coefficients show no comparable ordering, with a fitted gradient of $+0.010\pm0.028$. Thus, although the profile width responds substantially to luminosity for individual transitions, its luminosity response does not exhibit the systematic Balmer-order dependence seen in the central depths.

\begin{deluxetable*}{lcccccc}
\tabletypesize{\scriptsize}
\tablewidth{0pt}
\tablecaption{Direct Logarithmic Luminosity Responses of Balmer-Profile Observables}
\label{tab:profile_response}
\tablehead{
\colhead{Line} &
\colhead{$s_d$ (dex$^{-1}$)} &
\colhead{$s_{\rm FWHM}$ (dex$^{-1}$)} &
\colhead{$s_{EW}$ (dex$^{-1}$)} &
\colhead{$s_{\kappa}$ (dex$^{-1}$)} &
\colhead{95\% CI($s_{\kappa}$)} &
\colhead{$N$}
}
\startdata
H8          & $-0.108\pm0.030$ & $-0.456\pm0.076$ & $-0.550\pm0.115$ & $+0.014\pm0.067$ & $[-0.11,+0.14]$ & 45 \\
H$\epsilon$ & $-0.084\pm0.032$ & $-0.265\pm0.071$ & $-0.312\pm0.100$ & $+0.037\pm0.043$ & $[-0.05,+0.12]$ & 58 \\
H$\delta$   & $-0.124\pm0.027$ & $-0.574\pm0.087$ & $-0.703\pm0.108$ & $-0.005\pm0.041$ & $[-0.06,+0.06]$ & 61 \\
H$\gamma$   & $-0.200\pm0.037$ & $-0.529\pm0.082$ & $-0.711\pm0.109$ & $+0.017\pm0.033$ & $[-0.05,+0.08]$ & 72 \\
H$\beta$    & $-0.430\pm0.104$ & $-0.402\pm0.064$ & $-0.871\pm0.148$ & $-0.038\pm0.039$ & $[-0.16,+0.06]$ & 72 \\
H$\alpha$   & $-0.959\pm0.179$ & $-0.450\pm0.091$ & $-1.724\pm0.262$ & $-0.315\pm0.129$ & $[-0.59,+0.05]$ & 68
\enddata
\tablecomments{
$s_d$, $s_{\rm FWHM}$, and $s_{EW}$ denote the directly fitted logarithmic luminosity responses, $s_X=\partial\ln X/\partial\log(L/L_{\odot})$, at fixed $T_{\rm eff}$. The profile-shape coefficient $s_{\kappa}$ is obtained from a separate direct fit to $\ln\kappa$, where $\kappa=EW/(d\,{\rm FWHM})$. The identity $s_{\kappa}=s_{EW}-s_d-s_{\rm FWHM}$ is used only for interpretation and is not treated as a combination of statistically independent measurements. Quoted coefficient uncertainties are $1\sigma$ uncertainties, while the listed intervals for $s_{\kappa}$ are 95\% bootstrap intervals obtained by resampling whole stars.}
\end{deluxetable*}

As a sanity check on the construction of the shape parameter, an ideal Gaussian profile integrated over infinite limits has $\kappa=\sqrt{\pi/(4\ln2)}=1.0645$. For spectra for which MERLIS selects a single-Gaussian representation, the observed median is $\kappa=1.054$, consistent with the expected scale of the quantity. More generally, larger values of $\kappa$ correspond to profiles with a greater fraction of their absorption in the wings relative to the central depth and FWHM, so $\kappa$ provides an empirical measure of profile shape.

The directly fitted profile-shape response, $s_{\kappa}=\partial\ln\kappa/\partial\log(L/L_{\odot})$, is consistent with zero from H8 through H$\beta$. The corresponding bootstrap 95\% intervals all include zero, indicating that no systematic luminosity-correlated change in the dimensionless profile-shape factor is detected for these transitions.

H$\alpha$ behaves differently. The logarithmic profile-shape analysis contains 68 stars with positive central depth and equivalent width, for which the full-sample response is $s_{\kappa}=-0.315\pm0.129$, with a bootstrap 95\% interval of $[-0.59,+0.05]$. Of the seven objects flagged as having strongly emission-filled H$\alpha$ cores in the parent sample, only three enter this logarithmic sample; the remaining four have non-positive equivalent widths and therefore do not have a defined $\ln\kappa$. Excluding the three flagged objects that remain in the logarithmic sample reduces the H$\alpha$ subset from 68 to 65 stars and gives $s_{\kappa}=-0.394\pm0.063$, with a bootstrap 95\% interval of $[-0.525,-0.270]$. The most extreme emission-dominated profiles are therefore absent from the logarithmic $\kappa$ analysis because $\kappa$ is undefined when $EW\leq0$. The full-sample coefficient is not driven by these extreme cases, while removal of the three remaining filled-core profiles strengthens the negative shape response.

The H$\alpha$ result was additionally tested against the profile family selected by MERLIS. Because the preferred profile family is itself determined by the observed morphology, including the model label as an ordinary covariate may remove part of the shape dependence that the analysis is intended to measure. Restricting the analysis to the dominant H$\alpha$ profile family gives $s_{\kappa}=-0.174\pm0.064$, with a 95\% bootstrap interval of $[-0.270,-0.047]$, while including the model label as a covariate gives a negative point estimate of $-0.181$. Thus, the sign of the H$\alpha$ shape response is preserved, although its inferred amplitude depends on how profile-model selection is handled.

Repeated spectra of $\alpha$~Cyg provide an empirical check on this model dependence. The preferred H$\alpha$ model alternates between the double-Gaussian and Voigt-plus-Gaussian forms, seven epochs each, while the corresponding median $\kappa$ values are nearly identical, 1.499 and 1.501, with a total scatter of 5.5\%. Thus, in this multi-epoch case, the profile-model label follows the observed morphology rather than determining the measured $\kappa$. We therefore adopt the all-model result as the primary empirical estimate and regard the dominant-model-family result as a conservative sensitivity test.

Because $d$, FWHM, and $EW$ are measured from the same spectrum, their uncertainties are not independent. The bootstrap intervals quoted above preserve the observed joint variation of the three quantities across the stellar sample, but they do not constitute a reconstruction of the full per-spectrum measurement covariance. Repeated observations nevertheless demonstrate the expected dependence directly: in $\alpha$~Cyg, $\ln d$ and $\ln{\rm FWHM}$ are anticorrelated, with $r=-0.75$, $-0.76$, and $-0.80$ for H$\gamma$, H$\beta$, and H$\alpha$, respectively. The corresponding epoch-to-epoch scatter in $\ln\kappa$ is 0.033, 0.035, and 0.054. Because these repeated observations contain intrinsic stellar variability in addition to measurement noise, they provide an empirical upper bound on the repeatability scatter rather than an isolated measurement covariance.

\begin{figure*}[t]
\centering
\includegraphics[width=\textwidth]{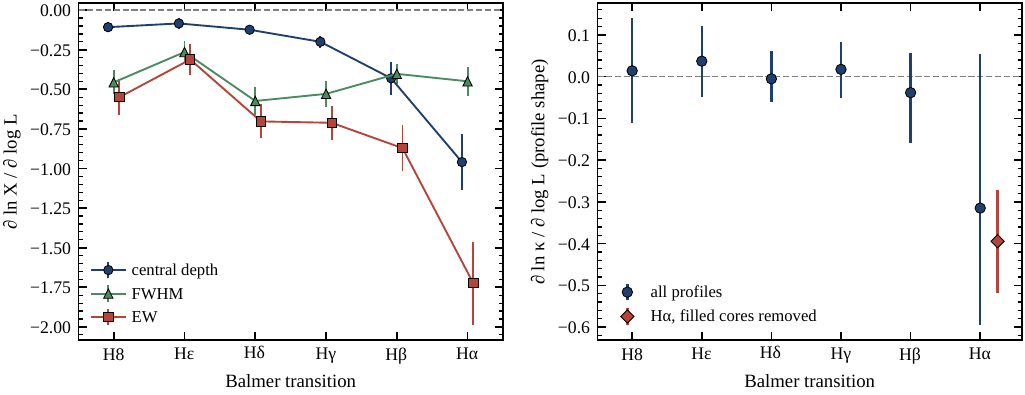}
\caption{
Direct logarithmic luminosity responses of the Balmer-profile observables.
\textit{Left:} Fitted responses $s_X=\partial\ln X/\partial\log(L/L_{\odot})$ at fixed $T_{\rm eff}$ for the central depth, FWHM, and equivalent width of each Balmer transition. The central-depth response becomes progressively stronger toward the lower-order members, whereas the FWHM response shows no comparable monotonic ordering.
\textit{Right:} Profile-shape response $s_{\kappa}=\partial\ln\kappa/\partial\log(L/L_{\odot})$, where $\kappa=EW/[d\,{\rm FWHM}]$. Error bars show the uncertainties of the inferred shape response. The $s_{\kappa}$ values are consistent with zero from H8 through H$\beta$. For the full logarithmic H$\alpha$ sample ($N=68$), the 95\% interval also includes zero; after excluding the three filled-core objects that remain in this sample ($N=65$), the shape response becomes significantly negative.}
\label{fig:profile_response}
\end{figure*}

\section{Discussion}
\label{sec:discussion}

\subsection{Balmer-order dependence and pressure broadening}
\label{sec:disc_stark}

The empirical sequence is qualitatively compatible with a systematic change in the importance of pressure broadening along the Balmer series. The scaling considered below is used only to illustrate the leading Balmer-order dependence of the characteristic Stark contribution; it is not a model of the measured luminosity coefficients. In particular, it does not predict the sign or amplitude of $q_n$, the central-depth response, or the line-formation depth of an individual transition. These quantities can depend simultaneously on surface gravity, effective temperature, atmospheric stratification, external broadening, non-LTE line formation, and winds.

Because hydrogen levels are degenerate in orbital angular momentum, hydrogen lines are subject to the linear Stark effect. Following the standard quasi-static treatment and Holtsmark microfield scaling \citep{1974slbp.book.....G}, the characteristic splitting of a level with principal quantum number $n$ in a microfield $F$ scales approximately as $\Delta E\propto n(n-1)e a_0F$. For a Balmer transition $n\rightarrow2$ with $n\ge3$, the upper-level contribution dominates to first order. Converting to wavelength units and using $F_0\propto n_{\rm e}^{2/3}$ gives

\begin{equation}
\Delta\lambda_{\rm Stark}
\propto
\lambda_n^{2}\,n(n-1)\,n_{\rm e}^{2/3},
\label{eq:stark_scaling}
\end{equation}

whereas the thermal Doppler width scales as $\Delta\lambda_{\rm D}\propto\lambda_n$. To isolate only the leading $n$- and wavelength-dependence of the relative Stark contribution, we define the dimensionless quantity

\begin{equation}
\mathcal{R}_n
\equiv
\frac{\lambda_n\,n(n-1)}
{\lambda_{{\rm H}\alpha}\times6},
\label{eq:R_n}
\end{equation}

which is normalized to unity for H$\alpha$ and increases to 5.53 for H8. $\mathcal{R}_n$ is intended only as a heuristic, order-of-magnitude descriptor. It is not a prediction of an observed line width, wing extent, central depth, or luminosity-response coefficient $q_n$.

This leading-order scaling does not replace a line-profile calculation. Detailed hydrogen Stark profiles require the unified-theory calculations of \citet{1973ApJS...25...37V} or the model-microfield calculations of \citet{1999A&AS..140...93S}, while neutral-hydrogen self-broadening of the Balmer wings is treated separately by \citet{2000A&A...363.1091B}.

The monotonic change in $\mathcal{R}_n$ therefore serves only to illustrate that the relative importance of pressure broadening is expected to vary systematically with Balmer order. Importantly, $\mathcal{R}_n$ and the measured luminosity coefficients are different quantities and are not expected to track one another directly: $\mathcal{R}_n$ increases toward the higher-order Balmer members, whereas the magnitude of the observed central-depth response increases toward the lower-order lines.

The empirical H8--H$\delta$ plateau followed by the stronger H$\gamma$--H$\alpha$ core response is therefore not presented as a direct consequence of Equation~(\ref{eq:stark_scaling}) or Equation~(\ref{eq:R_n}). Rather, the scaling provides a qualitative physical motivation for why different Balmer transitions may respond differently to changes in atmospheric density structure. Determining whether Stark broadening, gravity-dependent atmospheric stratification, or other line-formation effects quantitatively reproduce the measured $q_n$ sequence requires dedicated atmosphere calculations.

Accordingly, the fitted gradients $db/dn$ and $dq/dn$ should be regarded as compact empirical descriptors of the observed Balmer-order sequence, rather than as measurements of a fundamental $n$-dependent broadening law.

\subsection{H$\alpha$: possible non-LTE and wind contributions}
\label{sec:disc_halpha}

H$\alpha$ occupies the extreme low-order end of the measured sequence and is formed under conditions in which non-LTE and wind effects can be especially important. Detailed analyses of early-type stars show that accurate H$\alpha$ profiles require non-LTE treatment \citep{2004ApJ...609.1181P,2006A&A...445.1099P,2012A&A...543A..80F,2022A&A...668A..92W}, while the same transition is widely used as an optical mass-loss diagnostic because emission from the extended atmosphere and wind can fill the photospheric absorption profile \citep{2010A&A...521A...5C,2014A&A...565A..62P}.

The Balmer-order gradient remains positive when H$\alpha$ is excluded, so the sequence is not created by the wind-sensitive endpoint. The particularly strong H$\alpha$ coefficient can therefore be interpreted only as a possible additional non-LTE and/or wind-related contribution superimposed on the broader empirical progression. Quantitative separation of these effects requires dedicated atmosphere and wind calculations.

\subsection{Profile-shape response and equivalent width}
\label{sec:disc_ew}

The profile-shape analysis clarifies why equivalent width need not reproduce the same Balmer-order behaviour as the central depth. Because $EW$, $d$, and FWHM are measured from the same locally normalized profile, their responses are statistically coupled; the directly fitted $\kappa=EW/(d\,{\rm FWHM})$ term is therefore the appropriate empirical measure of any residual luminosity-correlated change in morphology.

From H8 through H$\beta$, $s_{\kappa}$ is consistent with zero, so the observed luminosity response can be described, within the present precision, through changes in depth and width without requiring a systematic change in the dimensionless profile-shape factor. H$\alpha$ is different: the full logarithmic sample is not formally significant at the 95\% level, but the shape response becomes significantly negative after the three remaining strongly filled-core profiles are removed. The model-family sensitivity tests preserve the negative sign but show that its amplitude is not yet uniquely determined. We therefore regard the H$\alpha$ shape result as evidence for an additional luminosity-correlated morphological change, with its quantitative magnitude still model dependent.

\subsection{Alternative interpretations and systematic effects}
\label{sec:disc_systematics}

External broadening is the clearest measured alternative contribution to the core-depth progression. The metal-line analysis shows that non-thermal broadening increases with luminosity, and both the direct covariate test and the convolution experiment indicate that this effect accounts for approximately 12--16\% of the measured differential gradient. It is therefore measurable but insufficient to reproduce the full sequence. The anticorrelation between ${\rm FWHM}({\rm H}\gamma)$ and the metal-line broadening diagnostic also confirms that a Balmer width cannot be treated as a clean proxy for rotation or macroturbulence at this resolution.

The remaining empirical checks point in the same direction. The result survives removal of flagged H$\epsilon$ and H$\alpha$ measurements, robust regression, restriction to the common six-line stellar subset, alternative temperature scales, and restriction to the post-2018 instrumental configuration. H8 blending by the nearly coincident He\,{\sc i} $\lambda3888.65$ transition and H$\epsilon$ contamination by Ca\,{\sc ii}~H remain line-specific limitations, but neither is required to produce the lower-order progression. Single-epoch variability is more important conceptually: the available VSX amplitudes increase with luminosity, and the multi-epoch spectra show that Balmer-core variability is especially strong toward H$\alpha$. We therefore treat variability as an additional luminosity-dependent source of scatter rather than as purely random, luminosity-independent noise.

\subsection{Limitations and outlook}
\label{sec:disc_outlook}

The principal limitation is that luminosity, rather than surface gravity, is the external predictor. Hydrogen-line formation responds directly to atmospheric density and gravity, while $L$, radius, and $g$ are coupled in evolved massive stars. The coefficients $q_n$ should therefore be interpreted as empirical luminosity-correlated responses. Repeating the analysis against spectroscopic or flux-weighted gravity,
$g_{\rm F}=g/T_{\rm eff}^4$ \citep{2003ApJ...582L..83K,2008ApJ...681..269K,2017AJ....154..102U},
would test whether the sequence is more compactly described as a gravity dependence.

A second limitation is the absence of a direct atmosphere-model comparison. Equation~(\ref{eq:stark_scaling}) describes a systematic variation in the relative importance of pressure broadening with Balmer order, but it does not predict the measured $q_n$ amplitudes. A dedicated grid spanning the observed $T_{\rm eff}$ and $\log g$ range, calculated for example with the hybrid approach of \citet{2006A&A...445.1099P}, \texttt{TLUSTY} \citep{1995ApJ...439..875H,2007ApJS..169...83L}, or \texttt{FASTWIND} \citep{2005A&A...435..669P}, degraded to the observational resolution and measured with the same pipeline, would provide the required quantitative test.

The third limitation is metallicity. The present sample is Galactic and spans a relatively narrow metallicity range, so the measured coefficients should not be transferred directly to lower-metallicity systems. Metallicity can alter atmospheric temperature and density structure, line blanketing, ionisation balance, and wind properties, while hydrogen itself may dominate the free-electron supply over part of the relevant temperature range. Magellanic Cloud supergiants provide a natural extension of the experiment \citep{2017AJ....154..102U}.

Finally, the present relations should not be regarded as a validated practical luminosity calibration. Any such application would require an independent validation sample, explicit resolution-degradation tests, and a quantitative treatment of contamination and observational precision, particularly for H$\alpha$.

\section{Conclusions}
\label{sec:conclusions}

We have measured the differential response of the optical Balmer series from H8 to H$\alpha$ in a programme sample of 84 Galactic B5--A5 luminous stars. Homogeneous TCO Balmer measurements are available for 80 stars, of which 72 form the primary luminosity-response sample. The principal conclusions are:

\begin{enumerate}

\item The Balmer central depths do not follow a single common scaling. Relative to H$\gamma$, the fitted line-to-line slopes form a monotonic sequence toward the lower-order members, with $db/dn=-0.129$ and a 95\% interval of $[-0.164,-0.095]$. The ordering is unchanged when H$\delta$ or H$\beta$ is used as the reference transition.

\item After controlling for $T_{\rm eff}$, the luminosity coefficients show an overall Balmer-order progression. The point-estimate gradient is $dq/dn=0.0419$, while the joint Monte Carlo analysis gives a median of 0.0387 and a 95\% interval of $[0.0253,0.0532]$. The H8--H$\delta$ coefficients form a statistically unresolved high-order plateau, followed by a well-supported increase in response magnitude toward H$\alpha$. The result survives the temperature-scale, sample-composition, flag-removal, robust-regression, instrumental-configuration, and individual-star sensitivity tests.

\item The luminosity response is not distributed uniformly within the profiles. The FWHM coefficients show no comparable Balmer-order progression, and the directly fitted profile-shape response is consistent with zero from H8 through H$\beta$. For H$\alpha$, the full logarithmic sample is not formally significant at the 95\% level, whereas removal of the three remaining strongly filled-core profiles yields a significantly negative shape response. External broadening contributes measurably but accounts for only approximately 12--16\% of the differential central-depth gradient.

\item The sequence is qualitatively compatible with the systematic variation of pressure broadening across the Balmer series, but luminosity is an empirical proxy rather than the fundamental line-formation variable. The enhanced H$\alpha$ response may contain additional non-LTE and wind contributions. Spectroscopic gravities, dedicated non-LTE atmosphere and wind models, and extension to lower-metallicity samples are required to identify the underlying physical dependence quantitatively.

\end{enumerate}

\section*{Data and Software Availability}
\phantomsection
\label{sec:avi}

The reduced continuum-normalized spectra and their associated metadata are publicly available through the Kazakhstan National Virtual Observatory (KazVO) via the \href{https://vo.fai.kz/obs_data.php?path=/tco_hot_supergiants/q/web}{KazVO spectral service}.

The complete machine-readable table used in the present analysis, including the adopted stellar parameters, Balmer-line measurements, uncertainties, and quality flags, is archived on Zenodo at \href{https://doi.org/10.5281/zenodo.23008571}{doi:10.5281/zenodo.23008571}.

The \textsc{MERLIS} source code, example configuration files, and line-list definitions used in this work are publicly available through the project repository at \url{https://github.com/nva1dman/merlis}.

\begin{acknowledgments}
This research was funded by the Science Committee of the Ministry of Science and Higher Education of the Republic of Kazakhstan (grant No. AP23484898). 
The authors are grateful to Anatoly S. Miroshnichenko for obtaining the TCO spectra used in this study and for his invaluable scientific guidance and support.
This research has made use of the SIMBAD database, operated at CDS, Strasbourg, France; SAO/NASA ADS, and Gaia data products.
\end{acknowledgments}

\software{
astropy \citep{2013A&A...558A..33A,2018AJ....156..123A,2022ApJ...935..167A},
SciPy \citep{2020NaMet..17..261V},
NumPy \citep{2020Natur.585..357H},
matplotlib \citep{2007CSE.....9...90H},
IRAF \citep{1986SPIE..627..733T,1993ASPC...52..173T}
}


\bibliography{sample701}{}

@ARTICLE{2006A&A...445.1099P,
       author = {{Przybilla}, N. and {Butler}, K. and {Becker}, S.~R. and {Kudritzki}, R.~P.},
        title = "{Quantitative spectroscopy of BA-type supergiants}",
      journal = {\aap},
         year = 2006,
        month = jan,
       volume = {445},
       number = {3},
        pages = {1099-1126},
          doi = {10.1051/0004-6361:20053832},
archivePrefix = {arXiv},
       eprint = {astro-ph/0509669},
 primaryClass = {astro-ph},
       adsurl = {https://ui.adsabs.harvard.edu/abs/2006A&A...445.1099P}
}

@ARTICLE{2012A&A...543A..80F,
       author = {{Firnstein}, M. and {Przybilla}, N.},
        title = "{Quantitative spectroscopy of Galactic BA-type supergiants. I. Atmospheric parameters}",
      journal = {\aap},
         year = 2012,
        month = jul,
       volume = {543},
          eid = {A80},
        pages = {A80},
          doi = {10.1051/0004-6361/201219034},
archivePrefix = {arXiv},
       eprint = {1207.0308},
 primaryClass = {astro-ph.SR},
       adsurl = {https://ui.adsabs.harvard.edu/abs/2012A&A...543A..80F}
}

@ARTICLE{2022A&A...668A..92W,
       author = {{We{\ss}mayer}, D. and {Przybilla}, N. and {Butler}, K.},
        title = "{Quantitative spectroscopy of B-type supergiants}",
      journal = {\aap},
         year = 2022,
        month = dec,
       volume = {668},
          eid = {A92},
        pages = {A92},
          doi = {10.1051/0004-6361/202243973},
archivePrefix = {arXiv},
       eprint = {2208.02692},
 primaryClass = {astro-ph.SR},
       adsurl = {https://ui.adsabs.harvard.edu/abs/2022A&A...668A..92W}
}

@ARTICLE{1964PDAO...12..317P,
       author = {{Petrie}, R.~M.},
        title = "{Spectroscopic Absolute Magnitudes of Early-Type Stars from Hydrogen Absorption-a Revision}",
      journal = {Publications of the Dominion Astrophysical Observatory Victoria},
         year = 1964,
        month = jan,
       volume = {12},
        pages = {317},
       adsurl = {https://ui.adsabs.harvard.edu/abs/1964PDAO...12..317P}
}

@ARTICLE{1974MNRAS.166..203B,
       author = {{Balona}, L. and {Crampton}, D.},
        title = "{The Hgamma -absolute magnitude calibration.}",
      journal = {\mnras},
         year = 1974,
        month = jan,
       volume = {166},
        pages = {203-217},
          doi = {10.1093/mnras/166.2.203},
       adsurl = {https://ui.adsabs.harvard.edu/abs/1974MNRAS.166..203B}
}

@ARTICLE{2004MNRAS.353..601E,
       author = {{Evans}, Christopher J. and {Howarth}, Ian D. and {Irwin}, Michael J. and {Burnley}, Adam W. and {Harries}, Timothy J.},
        title = "{A 2dF survey of the Small Magellanic Cloud}",
      journal = {\mnras},
         year = 2004,
        month = sep,
       volume = {353},
       number = {2},
        pages = {601-623},
          doi = {10.1111/j.1365-2966.2004.08096.x},
archivePrefix = {arXiv},
       eprint = {astro-ph/0406409},
 primaryClass = {astro-ph},
       adsurl = {https://ui.adsabs.harvard.edu/abs/2004MNRAS.353..601E}
}

@ARTICLE{2008ApJ...681..269K,
       author = {{Kudritzki}, Rolf-Peter and {Urbaneja}, Miguel A. and {Bresolin}, Fabio and {Przybilla}, Norbert and {Gieren}, Wolfgang and {Pietrzy{\'n}ski}, Grzegorz},
        title = "{Quantitative Spectroscopy of 24 A Supergiants in the Sculptor Galaxy NGC 300: Flux-weighted Gravity-Luminosity Relationship, Metallicity, and Metallicity Gradient}",
      journal = {\apj},
         year = 2008,
        month = jul,
       volume = {681},
       number = {1},
        pages = {269-289},
          doi = {10.1086/588647},
archivePrefix = {arXiv},
       eprint = {0803.3654},
 primaryClass = {astro-ph},
       adsurl = {https://ui.adsabs.harvard.edu/abs/2008ApJ...681..269K}
}

@ARTICLE{2017AJ....154..102U,
       author = {{Urbaneja}, M.~A. and {Kudritzki}, R.-P. and {Gieren}, W. and {Pietrzy{\'n}ski}, G. and {Bresolin}, F. and {Przybilla}, N.},
        title = "{LMC Blue Supergiant Stars and the Calibration of the Flux-weighted Gravity-Luminosity Relationship}",
      journal = {\aj},
         year = 2017,
        month = sep,
       volume = {154},
       number = {3},
          eid = {102},
        pages = {102},
          doi = {10.3847/1538-3881/aa79a8},
archivePrefix = {arXiv},
       eprint = {1706.03967},
 primaryClass = {astro-ph.SR},
       adsurl = {https://ui.adsabs.harvard.edu/abs/2017AJ....154..102U}
}

@ARTICLE{2023A&A...674A.212D,
       author = {{de Burgos}, A. and {Sim{\'o}n-D{\'\i}az}, S. and {Urbaneja}, M.~A. and {Negueruela}, I.},
        title = "{The IACOB project. IX. Building a modern empirical database of Galactic O9 - B9 supergiants: Sample selection, description, and completeness}",
      journal = {\aap},
         year = 2023,
        month = jun,
       volume = {674},
          eid = {A212},
        pages = {A212},
          doi = {10.1051/0004-6361/202346179},
archivePrefix = {arXiv},
       eprint = {2305.00305},
 primaryClass = {astro-ph.SR},
       adsurl = {https://ui.adsabs.harvard.edu/abs/2023A&A...674A.212D}
}

@ARTICLE{2024A&A...690A.176N,
       author = {{Negueruela}, I. and {Sim{\'o}n-D{\'\i}az}, S. and {de Burgos}, A. and {Casasbuenas}, A. and {Beck}, P.~G.},
        title = "{The IACOB project: XII. New grid of northern standards for the spectral classification of B-type stars}",
      journal = {\aap},
         year = 2024,
        month = oct,
       volume = {690},
          eid = {A176},
        pages = {A176},
          doi = {10.1051/0004-6361/202449298},
archivePrefix = {arXiv},
       eprint = {2407.04163},
 primaryClass = {astro-ph.SR},
       adsurl = {https://ui.adsabs.harvard.edu/abs/2024A&A...690A.176N}
}

@ARTICLE{2023Galax..11....8M,
       author = {{Miroshnichenko}, Anatoly S. and {Danford}, Stephen and {Andronov}, Ivan L. and {Aarnio}, Alicia N. and {Lauer}, Duncan and {Buroughs}, Holly},
        title = "{Refining Orbits of Bright Binary Systems}",
      journal = {Galaxies},
         year = 2023,
        month = jan,
       volume = {11},
       number = {1},
          eid = {8},
        pages = {8},
          doi = {10.3390/galaxies11010008},
       adsurl = {https://ui.adsabs.harvard.edu/abs/2023Galax..11....8M}
}

@ARTICLE{2025Univ...11..359V,
       author = {{Vaidman}, Nadezhda L. and {Nurmakhametova}, Shakhida T. and {Umirova}, Aziza B. and {Khokhlov}, Serik A. and {Agishev}, Aldiyar T. and {Yermekbayev}, Berik S.},
        title = "{Evaluating Gaia Astrometric Quality and Distances for Galactic Hot Supergiants}",
      journal = {Universe},
         year = 2025,
        month = oct,
       volume = {11},
       number = {11},
          eid = {359},
        pages = {359},
          doi = {10.3390/universe11110359},
       adsurl = {https://ui.adsabs.harvard.edu/abs/2025Univ...11..359V}
}

@ARTICLE{2010A&A...521A...5C,
       author = {{Chesneau}, O. and {Dessart}, L. and {Mourard}, D. and {B{\'e}rio}, Ph. and {Buil}, Ch. and {Bonneau}, D. and {Borges Fernandes}, M. and {Clausse}, J.~M. and {Delaa}, O. and {Marcotto}, A. and {Meilland}, A. and {Millour}, F. and {Nardetto}, N. and {Perraut}, K. and {Roussel}, A. and {Spang}, A. and {Stee}, P. and {Tallon-Bosc}, I. and {McAlister}, H. and {ten Brummelaar}, T. and {Sturmann}, J. and {Sturmann}, L. and {Turner}, N. and {Farrington}, C. and {Goldfinger}, P.~J.},
        title = "{Time, spatial, and spectral resolution of the H{\ensuremath{\alpha}} line-formation region of Deneb and Rigel with the VEGA/CHARA interferometer}",
      journal = {\aap},
         year = 2010,
        month = oct,
       volume = {521},
          eid = {A5},
        pages = {A5},
          doi = {10.1051/0004-6361/201014509},
archivePrefix = {arXiv},
       eprint = {1007.2095},
 primaryClass = {astro-ph.SR},
       adsurl = {https://ui.adsabs.harvard.edu/abs/2010A&A...521A...5C}
}

@ARTICLE{2014A&A...565A..62P,
       author = {{Petrov}, Blagovest and {Vink}, Jorick S. and {Gr{\"a}fener}, G{\"o}tz},
        title = "{On the H{\ensuremath{\alpha}} behaviour of blue supergiants: rise and fall over the bi-stability jump}",
      journal = {\aap},
         year = 2014,
        month = may,
       volume = {565},
          eid = {A62},
        pages = {A62},
          doi = {10.1051/0004-6361/201322754},
archivePrefix = {arXiv},
       eprint = {1403.4097},
 primaryClass = {astro-ph.SR},
       adsurl = {https://ui.adsabs.harvard.edu/abs/2014A&A...565A..62P}
}

@ARTICLE{1973ApJS...25...37V,
       author = {{Vidal}, C.~R. and {Cooper}, J. and {Smith}, E.~W.},
        title = "{Hydrogen Stark-Broadening Tables}",
      journal = {\apjs},
         year = 1973,
        month = jan,
       volume = {25},
        pages = {37},
          doi = {10.1086/190264},
       adsurl = {https://ui.adsabs.harvard.edu/abs/1973ApJS...25...37V}
}

@ARTICLE{1999A&AS..140...93S,
       author = {{Stehl{\'e}}, C. and {Hutcheon}, R.},
        title = "{Extensive tabulations of Stark broadened hydrogen line profiles}",
      journal = {\aaps},
         year = 1999,
        month = nov,
       volume = {140},
        pages = {93-97},
          doi = {10.1051/aas:1999118},
       adsurl = {https://ui.adsabs.harvard.edu/abs/1999A&AS..140...93S}
}

@ARTICLE{2004ApJ...609.1181P,
       author = {{Przybilla}, Norbert and {Butler}, Keith},
        title = "{Non-LTE Line Formation for Hydrogen Revisited}",
      journal = {\apj},
         year = 2004,
        month = jul,
       volume = {609},
       number = {2},
        pages = {1181-1191},
          doi = {10.1086/421316},
archivePrefix = {arXiv},
       eprint = {astro-ph/0406458},
 primaryClass = {astro-ph},
       adsurl = {https://ui.adsabs.harvard.edu/abs/2004ApJ...609.1181P}
}

@ARTICLE{2003ApJ...582L..83K,
       author = {{Kudritzki}, Rolf P. and {Bresolin}, Fabio and {Przybilla}, Norbert},
        title = "{A New Extragalactic Distance Determination Method Using the Flux-weighted Gravity of Late B and Early A Supergiants}",
      journal = {\apjl},
         year = 2003,
        month = jan,
       volume = {582},
       number = {2},
        pages = {L83-L86},
          doi = {10.1086/367690},
archivePrefix = {arXiv},
       eprint = {astro-ph/0212042},
 primaryClass = {astro-ph},
       adsurl = {https://ui.adsabs.harvard.edu/abs/2003ApJ...582L..83K}
}

@ARTICLE{1914ApJ....40..385A,
       author = {{Adams}, W.~S. and {Kohlsch{\"u}tter}, A.},
        title = "{Some spectral criteria for the determination of absolute stellar magnitudes.}",
      journal = {\apj},
         year = 1914,
        month = nov,
       volume = {40},
        pages = {385-398},
          doi = {10.1086/142132},
       adsurl = {https://ui.adsabs.harvard.edu/abs/1914ApJ....40..385A}
}

@ARTICLE{1935ApJ....81..187A,
       author = {{Adams}, Walter S. and {Joy}, Alfred H. and {Humason}, Milton L. and {Brayton}, Ada Margaret},
        title = "{The Spectroscopic Absolute Magnitudes and Parallaxes of 4179 Stars}",
      journal = {\apj},
         year = 1935,
        month = apr,
       volume = {81},
        pages = {187},
          doi = {10.1086/143628},
       adsurl = {https://ui.adsabs.harvard.edu/abs/1935ApJ....81..187A}
}

@BOOK{1943assw.book.....M,
       author = {{Morgan}, William Wilson and {Keenan}, Philip Childs and {Kellman}, Edith},
        title = "{An atlas of stellar spectra, with an outline of spectral classification}",
         year = 1943,
       adsurl = {https://ui.adsabs.harvard.edu/abs/1943assw.book.....M}
}

@ARTICLE{2007ApJS..169...83L,
       author = {{Lanz}, Thierry and {Hubeny}, Ivan},
        title = "{A Grid of NLTE Line-blanketed Model Atmospheres of Early B-Type Stars}",
      journal = {\apjs},
         year = 2007,
        month = mar,
       volume = {169},
       number = {1},
        pages = {83-104},
          doi = {10.1086/511270},
archivePrefix = {arXiv},
       eprint = {astro-ph/0611891},
 primaryClass = {astro-ph},
       adsurl = {https://ui.adsabs.harvard.edu/abs/2007ApJS..169...83L}
}

@ARTICLE{1995ApJ...439..875H,
       author = {{Hubeny}, I. and {Lanz}, T.},
        title = "{Non-LTE Line-blanketed Model Atmospheres of Hot Stars. I. Hybrid Complete Linearization/Accelerated Lambda Iteration Method}",
      journal = {\apj},
         year = 1995,
        month = feb,
       volume = {439},
        pages = {875},
          doi = {10.1086/175226},
       adsurl = {https://ui.adsabs.harvard.edu/abs/1995ApJ...439..875H}
}

@ARTICLE{2005A&A...435..669P,
       author = {{Puls}, J. and {Urbaneja}, M.~A. and {Venero}, R. and {Repolust}, T. and {Springmann}, U. and {Jokuthy}, A. and {Mokiem}, M.~R.},
        title = "{Atmospheric NLTE-models for the spectroscopic analysis of blue stars with winds. II. Line-blanketed models}",
      journal = {\aap},
         year = 2005,
        month = may,
       volume = {435},
       number = {2},
        pages = {669-698},
          doi = {10.1051/0004-6361:20042365},
archivePrefix = {arXiv},
       eprint = {astro-ph/0411398},
 primaryClass = {astro-ph},
       adsurl = {https://ui.adsabs.harvard.edu/abs/2005A&A...435..669P}
}

@article{2026ApJ...1008....4V,
  author  = {Vaidman, N. L. and Khokhlov, S. A. and Nurmakhametova, Sh. T. and Litvinov, N. A. and Agishev, A. T. and Khokhlov, A. A. and Agishev, D. T. and Alimbetova, D. A.},
  title   = {MERLIS: Model-driven Measurement of Spectral-line Parameters in Heterogeneous Stellar Spectra},
  journal = {The Astrophysical Journal},
  year    = {2026},
  volume  = {1008},
  number  = {1},
  pages   = {4},
  doi     = {10.3847/1538-4357/ae927f}
}

@ARTICLE{2013A&A...558A..33A,
       author = {{Astropy Collaboration} and {Robitaille}, Thomas P. and {Tollerud}, Erik J. and {Greenfield}, Perry and {Droettboom}, Michael and {Bray}, Erik and {Aldcroft}, Tom and {Davis}, Matt and {Ginsburg}, Adam and {Price-Whelan}, Adrian M. and {Kerzendorf}, Wolfgang E. and {Conley}, Alexander and {Crighton}, Neil and {Barbary}, Kyle and {Muna}, Demitri and {Ferguson}, Henry and {Grollier}, Fr{\'e}d{\'e}ric and {Parikh}, Madhura M. and {Nair}, Prasanth H. and {Unther}, Hans M. and {Deil}, Christoph and {Woillez}, Julien and {Conseil}, Simon and {Kramer}, Roban and {Turner}, James E.~H. and {Singer}, Leo and {Fox}, Ryan and {Weaver}, Benjamin A. and {Zabalza}, Victor and {Edwards}, Zachary I. and {Azalee Bostroem}, K. and {Burke}, D.~J. and {Casey}, Andrew R. and {Crawford}, Steven M. and {Dencheva}, Nadia and {Ely}, Justin and {Jenness}, Tim and {Labrie}, Kathleen and {Lim}, Pey Lian and {Pierfederici}, Francesco and {Pontzen}, Andrew and {Ptak}, Andy and {Refsdal}, Brian and {Servillat}, Mathieu and {Streicher}, Ole},
        title = "{Astropy: A community Python package for astronomy}",
      journal = {\aap},
         year = 2013,
        month = oct,
       volume = {558},
          eid = {A33},
        pages = {A33},
          doi = {10.1051/0004-6361/201322068},
archivePrefix = {arXiv},
       eprint = {1307.6212},
 primaryClass = {astro-ph.IM},
       adsurl = {https://ui.adsabs.harvard.edu/abs/2013A&A...558A..33A}
}

@ARTICLE{2018AJ....156..123A,
       author = {{Astropy Collaboration} and {Price-Whelan}, A.~M. and {Sip{\H{o}}cz}, B.~M. and {G{\"u}nther}, H.~M. and {Lim}, P.~L. and {Crawford}, S.~M. and {Conseil}, S. and {Shupe}, D.~L. and {Craig}, M.~W. and {Dencheva}, N. and {Ginsburg}, A. and {VanderPlas}, J.~T. and {Bradley}, L.~D. and {P{\'e}rez-Su{\'a}rez}, D. and {de Val-Borro}, M. and {Aldcroft}, T.~L. and {Cruz}, K.~L. and {Robitaille}, T.~P. and {Tollerud}, E.~J. and {Ardelean}, C. and {Babej}, T. and {Bach}, Y.~P. and {Bachetti}, M. and {Bakanov}, A.~V. and {Bamford}, S.~P. and {Barentsen}, G. and {Barmby}, P. and {Baumbach}, A. and {Berry}, K.~L. and {Biscani}, F. and {Boquien}, M. and {Bostroem}, K.~A. and {Bouma}, L.~G. and {Brammer}, G.~B. and {Bray}, E.~M. and {Breytenbach}, H. and {Buddelmeijer}, H. and {Burke}, D.~J. and {Calderone}, G. and {Cano Rodr{\'\i}guez}, J.~L. and {Cara}, M. and {Cardoso}, J.~V.~M. and {Cheedella}, S. and {Copin}, Y. and {Corrales}, L. and {Crichton}, D. and {D'Avella}, D. and {Deil}, C. and {Depagne}, {\'E}. and {Dietrich}, J.~P. and {Donath}, A. and {Droettboom}, M. and {Earl}, N. and {Erben}, T. and {Fabbro}, S. and {Ferreira}, L.~A. and {Finethy}, T. and {Fox}, R.~T. and {Garrison}, L.~H. and {Gibbons}, S.~L.~J. and {Goldstein}, D.~A. and {Gommers}, R. and {Greco}, J.~P. and {Greenfield}, P. and {Groener}, A.~M. and {Grollier}, F. and {Hagen}, A. and {Hirst}, P. and {Homeier}, D. and {Horton}, A.~J. and {Hosseinzadeh}, G. and {Hu}, L. and {Hunkeler}, J.~S. and {Ivezi{\'c}}, {\v{Z}}. and {Jain}, A. and {Jenness}, T. and {Kanarek}, G. and {Kendrew}, S. and {Kern}, N.~S. and {Kerzendorf}, W.~E. and {Khvalko}, A. and {King}, J. and {Kirkby}, D. and {Kulkarni}, A.~M. and {Kumar}, A. and {Lee}, A. and {Lenz}, D. and {Littlefair}, S.~P. and {Ma}, Z. and {Macleod}, D.~M. and {Mastropietro}, M. and {McCully}, C. and {Montagnac}, S. and {Morris}, B.~M. and {Mueller}, M. and {Mumford}, S.~J. and {Muna}, D. and {Murphy}, N.~A. and {Nelson}, S. and {Nguyen}, G.~H. and {Ninan}, J.~P. and {N{\"o}the}, M. and {Ogaz}, S. and {Oh}, S. and {Parejko}, J.~K. and {Parley}, N. and {Pascual}, S. and {Patil}, R. and {Patil}, A.~A. and {Plunkett}, A.~L. and {Prochaska}, J.~X. and {Rastogi}, T. and {Reddy Janga}, V. and {Sabater}, J. and {Sakurikar}, P. and {Seifert}, M. and {Sherbert}, L.~E. and {Sherwood-Taylor}, H. and {Shih}, A.~Y. and {Sick}, J. and {Silbiger}, M.~T. and {Singanamalla}, S. and {Singer}, L.~P. and {Sladen}, P.~H. and {Sooley}, K.~A. and {Sornarajah}, S. and {Streicher}, O. and {Teuben}, P. and {Thomas}, S.~W. and {Tremblay}, G.~R. and {Turner}, J.~E.~H. and {Terr{\'o}n}, V. and {van Kerkwijk}, M.~H. and {de la Vega}, A. and {Watkins}, L.~L. and {Weaver}, B.~A. and {Whitmore}, J.~B. and {Woillez}, J. and {Zabalza}, V. and {Astropy Contributors}},
        title = "{The Astropy Project: Building an Open-science Project and Status of the v2.0 Core Package}",
      journal = {\aj},
         year = 2018,
        month = sep,
       volume = {156},
       number = {3},
          eid = {123},
        pages = {123},
          doi = {10.3847/1538-3881/aabc4f},
archivePrefix = {arXiv},
       eprint = {1801.02634},
 primaryClass = {astro-ph.IM},
       adsurl = {https://ui.adsabs.harvard.edu/abs/2018AJ....156..123A}
}

@ARTICLE{2022ApJ...935..167A,
       author = {{Astropy Collaboration} and {Price-Whelan}, Adrian M. and {Lim}, Pey Lian and {Earl}, Nicholas and {Starkman}, Nathaniel and {Bradley}, Larry and {Shupe}, David L. and {Patil}, Aarya A. and {Corrales}, Lia and {Brasseur}, C.~E. and {N{\"o}the}, Maximilian and {Donath}, Axel and {Tollerud}, Erik and {Morris}, Brett M. and {Ginsburg}, Adam and {Vaher}, Eero and {Weaver}, Benjamin A. and {Tocknell}, James and {Jamieson}, William and {van Kerkwijk}, Marten H. and {Robitaille}, Thomas P. and {Merry}, Bruce and {Bachetti}, Matteo and {G{\"u}nther}, H. Moritz and {Aldcroft}, Thomas L. and {Alvarado-Montes}, Jaime A. and {Archibald}, Anne M. and {B{\'o}di}, Attila and {Bapat}, Shreyas and {Barentsen}, Geert and {Baz{\'a}n}, Juanjo and {Biswas}, Manish and {Boquien}, M{\'e}d{\'e}ric and {Burke}, D.~J. and {Cara}, Daria and {Cara}, Mihai and {Conroy}, Kyle E. and {Conseil}, Simon and {Craig}, Matthew W. and {Cross}, Robert M. and {Cruz}, Kelle L. and {D'Eugenio}, Francesco and {Dencheva}, Nadia and {Devillepoix}, Hadrien A.~R. and {Dietrich}, J{\"o}rg P. and {Eigenbrot}, Arthur Davis and {Erben}, Thomas and {Ferreira}, Leonardo and {Foreman-Mackey}, Daniel and {Fox}, Ryan and {Freij}, Nabil and {Garg}, Suyog and {Geda}, Robel and {Glattly}, Lauren and {Gondhalekar}, Yash and {Gordon}, Karl D. and {Grant}, David and {Greenfield}, Perry and {Groener}, Austen M. and {Guest}, Steve and {Gurovich}, Sebastian and {Handberg}, Rasmus and {Hart}, Akeem and {Hatfield-Dodds}, Zac and {Homeier}, Derek and {Hosseinzadeh}, Griffin and {Jenness}, Tim and {Jones}, Craig K. and {Joseph}, Prajwel and {Kalmbach}, J. Bryce and {Karamehmetoglu}, Emir and {Ka{\l}uszy{\'n}ski}, Miko{\l}aj and {Kelley}, Michael S.~P. and {Kern}, Nicholas and {Kerzendorf}, Wolfgang E. and {Koch}, Eric W. and {Kulumani}, Shankar and {Lee}, Antony and {Ly}, Chun and {Ma}, Zhiyuan and {MacBride}, Conor and {Maljaars}, Jakob M. and {Muna}, Demitri and {Murphy}, N.~A. and {Norman}, Henrik and {O'Steen}, Richard and {Oman}, Kyle A. and {Pacifici}, Camilla and {Pascual}, Sergio and {Pascual-Granado}, J. and {Patil}, Rohit R. and {Perren}, Gabriel I. and {Pickering}, Timothy E. and {Rastogi}, Tanuj and {Roulston}, Benjamin R. and {Ryan}, Daniel F. and {Rykoff}, Eli S. and {Sabater}, Jose and {Sakurikar}, Parikshit and {Salgado}, Jes{\'u}s and {Sanghi}, Aniket and {Saunders}, Nicholas and {Savchenko}, Volodymyr and {Schwardt}, Ludwig and {Seifert-Eckert}, Michael and {Shih}, Albert Y. and {Jain}, Anany Shrey and {Shukla}, Gyanendra and {Sick}, Jonathan and {Simpson}, Chris and {Singanamalla}, Sudheesh and {Singer}, Leo P. and {Singhal}, Jaladh and {Sinha}, Manodeep and {Sip{\H{o}}cz}, Brigitta M. and {Spitler}, Lee R. and {Stansby}, David and {Streicher}, Ole and {{\v{S}}umak}, Jani and {Swinbank}, John D. and {Taranu}, Dan S. and {Tewary}, Nikita and {Tremblay}, Grant R. and {de Val-Borro}, Miguel and {Van Kooten}, Samuel J. and {Vasovi{\'c}}, Zlatan and {Verma}, Shresth and {de Miranda Cardoso}, Jos{\'e} Vin{\'\i}cius and {Williams}, Peter K.~G. and {Wilson}, Tom J. and {Winkel}, Benjamin and {Wood-Vasey}, W.~M. and {Xue}, Rui and {Yoachim}, Peter and {Zhang}, Chen and {Zonca}, Andrea and {Astropy Project Contributors}},
        title = "{The Astropy Project: Sustaining and Growing a Community-oriented Open-source Project and the Latest Major Release (v5.0) of the Core Package}",
      journal = {\apj},
         year = 2022,
        month = aug,
       volume = {935},
       number = {2},
          eid = {167},
        pages = {167},
          doi = {10.3847/1538-4357/ac7c74},
archivePrefix = {arXiv},
       eprint = {2206.14220},
 primaryClass = {astro-ph.IM},
       adsurl = {https://ui.adsabs.harvard.edu/abs/2022ApJ...935..167A}
}

@ARTICLE{2020NaMet..17..261V,
       author = {{Virtanen}, Pauli and {Gommers}, Ralf and {Oliphant}, Travis E. and {Haberland}, Matt and {Reddy}, Tyler and {Cournapeau}, David and {Burovski}, Evgeni and {Peterson}, Pearu and {Weckesser}, Warren and {Bright}, Jonathan and {van der Walt}, St{\'e}fan J. and {Brett}, Matthew and {Wilson}, Joshua and {Millman}, K. Jarrod and {Mayorov}, Nikolay and {Nelson}, Andrew R.~J. and {Jones}, Eric and {Kern}, Robert and {Larson}, Eric and {Carey}, C.~J. and {Polat}, {\.I}lhan and {Feng}, Yu and {Moore}, Eric W. and {VanderPlas}, Jake and {Laxalde}, Denis and {Perktold}, Josef and {Cimrman}, Robert and {Henriksen}, Ian and {Quintero}, E.~A. and {Harris}, Charles R. and {Archibald}, Anne M. and {Ribeiro}, Ant{\^o}nio H. and {Pedregosa}, Fabian and {van Mulbregt}, Paul and {SciPy 1.  0 Contributors}},
        title = "{SciPy 1.0: fundamental algorithms for scientific computing in Python}",
      journal = {Nature Methods},
         year = 2020,
        month = feb,
       volume = {17},
        pages = {261-272},
          doi = {10.1038/s41592-019-0686-2},
archivePrefix = {arXiv},
       eprint = {1907.10121},
 primaryClass = {cs.MS},
       adsurl = {https://ui.adsabs.harvard.edu/abs/2020NaMet..17..261V}
}

@ARTICLE{2020Natur.585..357H,
       author = {{Harris}, Charles R. and {Millman}, K. Jarrod and {van der Walt}, St{\'e}fan J. and {Gommers}, Ralf and {Virtanen}, Pauli and {Cournapeau}, David and {Wieser}, Eric and {Taylor}, Julian and {Berg}, Sebastian and {Smith}, Nathaniel J. and {Kern}, Robert and {Picus}, Matti and {Hoyer}, Stephan and {van Kerkwijk}, Marten H. and {Brett}, Matthew and {Haldane}, Allan and {del R{\'\i}o}, Jaime Fern{\'a}ndez and {Wiebe}, Mark and {Peterson}, Pearu and {G{\'e}rard-Marchant}, Pierre and {Sheppard}, Kevin and {Reddy}, Tyler and {Weckesser}, Warren and {Abbasi}, Hameer and {Gohlke}, Christoph and {Oliphant}, Travis E.},
        title = "{Array programming with NumPy}",
      journal = {\nat},
         year = 2020,
        month = sep,
       volume = {585},
       number = {7825},
        pages = {357-362},
          doi = {10.1038/s41586-020-2649-2},
archivePrefix = {arXiv},
       eprint = {2006.10256},
 primaryClass = {cs.MS},
       adsurl = {https://ui.adsabs.harvard.edu/abs/2020Natur.585..357H}
}

@ARTICLE{2007CSE.....9...90H,
       author = {{Hunter}, John D.},
        title = "{Matplotlib: A 2D Graphics Environment}",
      journal = {Computing in Science and Engineering},
         year = 2007,
        month = jan,
       volume = {9},
       number = {3},
        pages = {90-95},
          doi = {10.1109/MCSE.2007.55},
       adsurl = {https://ui.adsabs.harvard.edu/abs/2007CSE.....9...90H}
}

@INPROCEEDINGS{1986SPIE..627..733T,
       author = {{Tody}, Doug},
        title = "{The IRAF Data Reduction and Analysis System}",
    booktitle = {Instrumentation in astronomy VI},
         year = 1986,
       editor = {{Crawford}, David L.},
       series = {Society of Photo-Optical Instrumentation Engineers (SPIE) Conference Series},
       volume = {627},
        month = jan,
        pages = {733},
          doi = {10.1117/12.968154},
       adsurl = {https://ui.adsabs.harvard.edu/abs/1986SPIE..627..733T}
}

@INPROCEEDINGS{1993ASPC...52..173T,
       author = {{Tody}, Doug},
        title = "{IRAF in the Nineties}",
    booktitle = {Astronomical Data Analysis Software and Systems II},
         year = 1993,
       editor = {{Hanisch}, R.~J. and {Brissenden}, R.~J.~V. and {Barnes}, J.},
       series = {Astronomical Society of the Pacific Conference Series},
       volume = {52},
        month = jan,
        pages = {173},
       adsurl = {https://ui.adsabs.harvard.edu/abs/1993ASPC...52..173T}
}

@ARTICLE{2012A&A...537A.146E,
       author = {{Ekstr{\"o}m}, S. and {Georgy}, C. and {Eggenberger}, P. and {Meynet}, G. and {Mowlavi}, N. and {Wyttenbach}, A. and {Granada}, A. and {Decressin}, T. and {Hirschi}, R. and {Frischknecht}, U. and {Charbonnel}, C. and {Maeder}, A.},
        title = "{Grids of stellar models with rotation. I. Models from 0.8 to 120 M$_{☉}$ at solar metallicity (Z = 0.014)}",
      journal = {\aap},
         year = 2012,
        month = jan,
       volume = {537},
          eid = {A146},
        pages = {A146},
          doi = {10.1051/0004-6361/201117751},
archivePrefix = {arXiv},
       eprint = {1110.5049},
 primaryClass = {astro-ph.SR},
       adsurl = {https://ui.adsabs.harvard.edu/abs/2012A&A...537A.146E}
}

@Article{galaxies14050087,
AUTHOR = {Nurmakhametova, Shakhida T. and Umirova, Aziza B. and Vaidman, Nadezhda L. and Khokhlov, Serik A. and Agishev, Aldiyar T. and Agishev, Damir T.},
TITLE = {An Observational Hertzsprung–Russell Diagram and Fundamental Parameters of Galactic BA Supergiants},
JOURNAL = {Galaxies},
VOLUME = {14},
YEAR = {2026},
NUMBER = {5},
ARTICLE-NUMBER = {87},
URL = {https://www.mdpi.com/2075-4434/14/5/87},
ISSN = {2075-4434},
DOI = {10.3390/galaxies14050087}
}

@ARTICLE{2019ApJ...887...93G,
       author = {{Green}, Gregory M. and {Schlafly}, Edward and {Zucker}, Catherine and {Speagle}, Joshua S. and {Finkbeiner}, Douglas},
        title = "{A 3D Dust Map Based on Gaia, Pan-STARRS 1, and 2MASS}",
      journal = {\apj},
         year = 2019,
        month = dec,
       volume = {887},
       number = {1},
          eid = {93},
        pages = {93},
          doi = {10.3847/1538-4357/ab5362},
archivePrefix = {arXiv},
       eprint = {1905.02734},
 primaryClass = {astro-ph.GA},
       adsurl = {https://ui.adsabs.harvard.edu/abs/2019ApJ...887...93G}
}

@ARTICLE{1989ApJ...345..245C,
       author = {{Cardelli}, Jason A. and {Clayton}, Geoffrey C. and {Mathis}, John S.},
        title = "{The Relationship between Infrared, Optical, and Ultraviolet Extinction}",
      journal = {\apj},
         year = 1989,
        month = oct,
       volume = {345},
        pages = {245},
          doi = {10.1086/167900},
       adsurl = {https://ui.adsabs.harvard.edu/abs/1989ApJ...345..245C}
}

@ARTICLE{1998A&A...333..231B,
       author = {{Bessell}, M.~S. and {Castelli}, F. and {Plez}, B.},
        title = "{Model atmospheres broad-band colors, bolometric corrections and temperature calibrations for O - M stars}",
      journal = {\aap},
         year = 1998,
        month = may,
       volume = {333},
        pages = {231-250},
       adsurl = {https://ui.adsabs.harvard.edu/abs/1998A&A...333..231B}
}

@misc{vaidman2026empiricaleffectivetemperaturecalibrationsgalactic,
      title={An Empirical Effective-Temperature Calibrations for Galactic B/A Supergiants}, 
      author={N. L. Vaidman and A. S. Miroshnichenko and Sh. T. Nurmakhametova and S. Danford and S. A. Khokhlov and I. M. Izmailova and Ch. T. Omarov and S. V. Zharikov and D. E. Yakhiya and A. T. Agishev and A. A. Khokhlov and D. T. Agishev},
      year={2026},
      eprint={2608.18619},
      archivePrefix={arXiv},
      primaryClass={astro-ph.SR},
      url={https://arxiv.org/abs/2608.18619}, 
}

@BOOK{1974slbp.book.....G,
       author = {{Griem}, H.~R.},
        title = "{Spectral line broadening by plasmas}",
         year = 1974,
       adsurl = {https://ui.adsabs.harvard.edu/abs/1974slbp.book.....G}
}

@ARTICLE{2000A&A...363.1091B,
       author = {{Barklem}, P.~S. and {Piskunov}, N. and {O'Mara}, B.~J.},
        title = "{Self-broadening in Balmer line wing formation in stellar atmospheres}",
      journal = {\aap},
         year = 2000,
        month = nov,
       volume = {363},
        pages = {1091-1105},
          doi = {10.48550/arXiv.astro-ph/0010022},
archivePrefix = {arXiv},
       eprint = {astro-ph/0010022},
 primaryClass = {astro-ph},
       adsurl = {https://ui.adsabs.harvard.edu/abs/2000A&A...363.1091B}
}
\bibliographystyle{aasjournalv7}

\end{document}